\documentclass[preprint,12pt,3p]{elsarticle}

\usepackage{amssymb}
\usepackage{multirow}

\usepackage{doi}

\usepackage{caption} 
\usepackage{graphicx,subfigure}

\usepackage{lineno}

\journal{Nucl. Instrum. Methods. Phys. Res. A}

\begin{document}

\begin{frontmatter}

\title{ Simulation and Efficiency Studies of Optical Photon Transportation and Detection with Plastic Antineutrino Detector Modules }
%\tnotetext[label0]{This is only an example}

\author[label1,label2]{Mustafa Kandemir\corref{cor1}}

\address[label1]{Department of Physics Eng., Istanbul Technical University, 34469, Istanbul, Turkey}
\address[label2]{Department of Physics, Recep Tayyip Erdogan University, 53100, Rize, Turkey\fnref{label4}}

\cortext[cor1]{corresponding author}
%\fntext[label3]{I also want to inform about\ldots}
%\fntext[label4]{Small city}

\ead{mustafa.kandemir@erdogan.edu.tr}

%\ead[url]{author-one-homepage.com}

\author[label1]{Altan Cakir}
\ead{cakir@cern.ch}

%\author[label1,label5]{Author Three}
%\ead{author.three@mail.com}

\begin{abstract}

In this work, the simulation of optical photons is carried out in an antineutrino detector module consisting of a plastic scintillator connected to light guides and photomultipliers on both ends, which is considered to be used for remote reactor monitoring in the field of nuclear safety. Using Monte Carlo (MC) based GEANT4 simulation, numerous parameters influencing the light collection and thereby the energy resolution of the antineutrino detector module are studied: e.g., degrees of scintillator surface roughness, reflector type, and its applying method onto scintillator and light guide surface, the reflectivity of the reflector, light guide geometries and diameter of the photocathode. The impact of each parameter is investigated by looking at the detected spectrum, i.e. the number photoelectrons per depositing energy. In addition, the average light collection efficiency of the detector module and its spatial variation are calculated for each simulation setup. According to the simulation results, it is found that photocathode size, light guide shape, reflectivity of reflecting material and wrapping method show a significant impact on the light collection efficiency while scintillator surface polishing level and the choose of reflector type show relatively less impact. This study demonstrates that these parameters are very important in the design of plastic scintillator included antineutrino detectors to improve the energy resolution efficiency.   

\end{abstract}

\begin{keyword}
Antineutrino \sep GEANT4  \sep  Optical photon \sep Light collection efficiency
\end{keyword}

\end{frontmatter}

 %\linenumbers

\section{Introduction}
\label{sec1}

Since the idea of monitoring nuclear reactors with antineutrinos first introduced in 1978 \cite{Mikaelyan}, many experiments have been conducted around the world \cite{Yuvv,Bernstein,Pequignot,Anjos}. The method used in these experiments is based on measuring the energy or number of antineutrinos emitted by reactors. Due to the fact that the number and energy spectra of the emitted antineutrinos per fission changes with respect to fission isotopes ($^{235}$U, $^{238}$U, $^{239}$Pu and $^{241}$Pu), and the relative contributions to fission of these isotopes evolve throughout the reactor cycle, measurement of the antineutrino energy spectrum over the fuel cycle provides information about fissile content of the reactor core.  

It is well known that antineutrinos are commonly detected by liquid-based scintillation detectors. However, these detectors should be relatively compact in size and preferably movable considering in the context of nuclear safety matters \cite{ABernstein}. Additionally, liquid scintillation detectors composed of hazardous substances and could be harmful during the transportation process because of containing flammable substances. Deploying this type of detector near nuclear reactors would be difficult. Cherenkov detectors are another opportunity for antineutrino detection. It is neither toxic nor flammable as liquid scintillation detectors. But the disadvantage of these detectors is that they produce very low light output and thereby less energy resolution compared to scintillation detectors. Moreover, antineutrinos from nuclear reactors tend to have energies around 2-3 MeV and this is too low for water-based Cherenkov detectors. Therefore, an alternative approach is to use plastic scintillation detectors.

%\appendix

 There are many proposals for the monitoring of nuclear reactors with plastic scintillation detectors \cite{Oguri,Battaglieri,Alekseev}. Suggested antineutrino detectors in these studies consist of identical detector modules. Each module has a plastic bar connected to light guides and photomultiplier tubes on both ends. Such a detector type to detect antineutrinos aboveground, it should have relatively strong background rejection and good energy resolution to measure antineutrino spectrum precisely. While 15$\%$ FWHM at 1 MeV energy resolution is enough to measure fuel composition evolution with plastic antineutrino detectors \cite{Georgadze}, achieving a higher energy resolution provides more accurate spectral measurement and increases the precision of spectral analysis and consequently enhances sensitivity level of the detector to the changes of isotopic composition of the reactor core. In addition, better energy resolution ensures better background rejection. This open a new area of further development to improve energy resolution of plastic detectors \cite{ABernstein}. Although this study consider as physics case the typical experimental arrangement of reactor's antineutrino detectors, the obtained results are applicable to any set up making use of plastic scintillator-light guide-PMTs trilogy.
 
The best energy resolution is obtained only by collecting the maximum number of photons. For this reason a very good light collection is necessary, so the parameters affecting light collection are an important criterion. For instance, the choice of reflector type and its applying method onto surfaces, reflectivity coefficient of wrapping material, reflection and refraction of the photon at boundary media, degree of scintillator surface roughness, shape of the light guide and photocathode effective area all have a significant impact on the light collection and thereby the energy resolution of the detector. Our goal is to explain the effect of each of these parameters efficiencies taking advantage of GEANT4 \cite{Agostinelli} simulation for optical photon transportation and detection in plastics scintillators. It is the first analysis to account for the detail simulation of scintillation photon in the context of antineutrino detection with plastic scintillator.

Geant4 is a toolkit for the simulation of the passage of particles through matter. In relation to the simulation of particles inside a scintillation detector, it has the capability to simulate both ionizing particles and scintillation light simultaneously. The simulation starts with the entrance of a charged particle into the detector and ends with the detection of the ensuing scintillation photons on photosensitive areas. All the physics processes from the initial energy deposition to the output signal are simulated.

 When a charged particle deposits its energy in the scintillator, optical photons are emitted isotropically with random polarization. These emitted photons may undergo three kinds of interactions depending on the material and surface properties: Elastic scattering, absorption, and medium boundary interactions. In an optical simulation, the user defines all material and surface properties of the detector to be used. Material properties inspect the generation of optical photons (emission spectrum, scintillation yield, etc.) and their transport through matter (elastic scattering and absorption). Surface properties inspect the physics processes when an optical photon arrives at a boundary between two media. For boundary process, there are two optical simulation models. The UNIFIED \cite{Levin} model and the GLISUR model. We are going to select the UNIFIED model, which is more comprehensive than the GLISUR model in terms of the surface wrappings.

\begin{figure}[!htb]
\centering
\begin{minipage}[t]{.47\textwidth}
  \centering
  \includegraphics[width=1.0\linewidth]{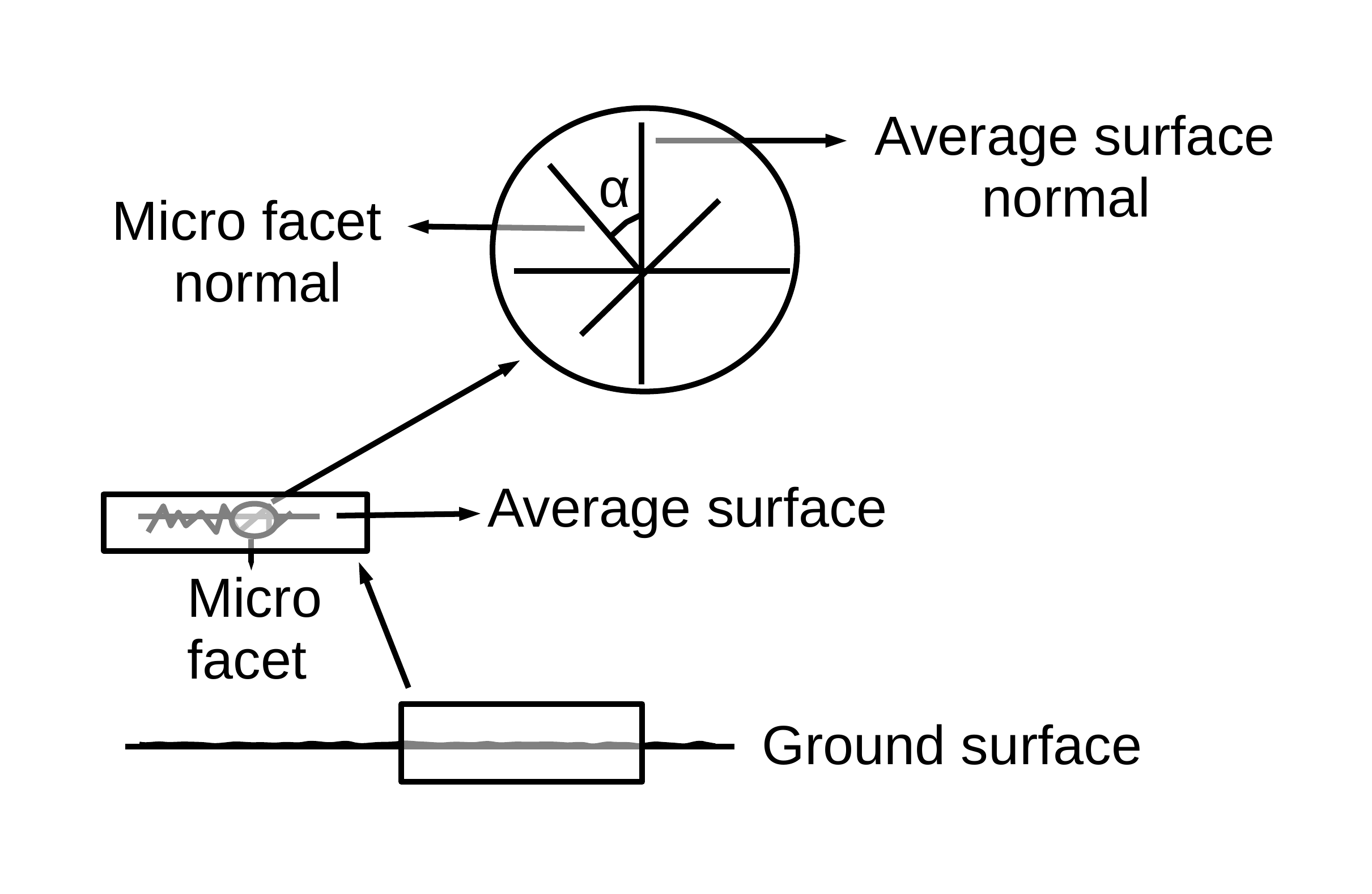}
 \caption{A ground surface made up of micro-facets. $\alpha$ is the angle between the average surface normal and micro-facet normal. $\sigma_{\alpha}$ is the standard deviation of this Gaussian distributed angle.   }
\label{fig:ground}
\end{minipage}%
\quad \quad
\begin{minipage}[t]{.47\textwidth}
  \centering
  \includegraphics[width=1.0\linewidth]{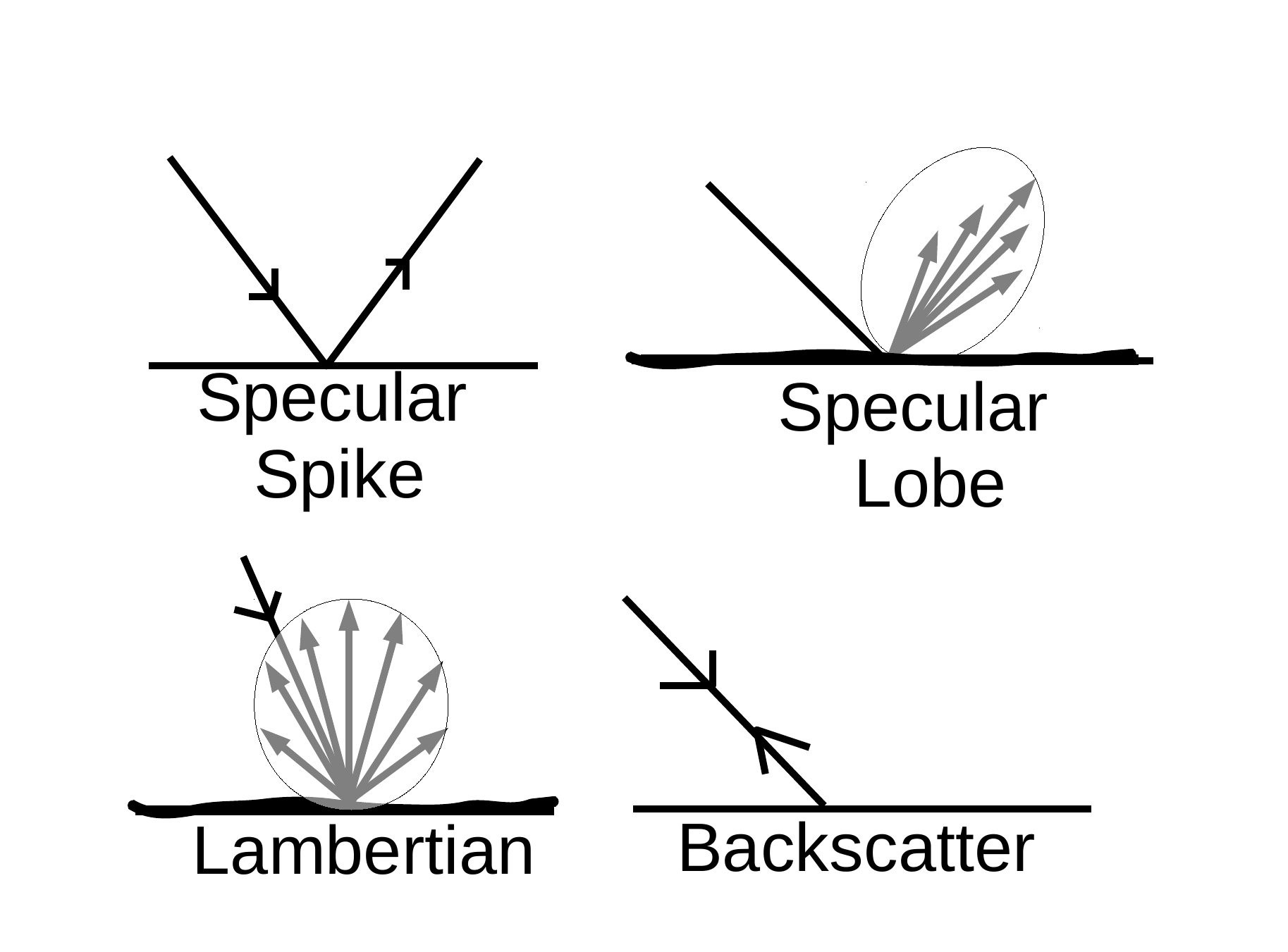}
  \caption{ Reflection types in UNIFIED model. Length of the arrow in circle indicates radiant flux at different reflection angles. }
\label{fig:reflection}
\end{minipage}
\end{figure}

The UNIFIED model applies to dielectric-dielectric interfaces and provides a realistic simulation model overcomes all aspects of surface finish and reflector coating. The surface between two media could be polished and ground. When a photon encounters polished surface between two media, Snell's law is applied based on the refractive index of two media. On the other hand, if the surface encountered is ground firstly a micro facet selected according to a parameter sigma alpha ($\sigma_{\alpha}$), which is the standard deviation of the angle between average surface normal and micro-facet normal (See Fig. \ref{fig:ground}), and then Snell's law is applied with respect to this facet normal. If Snell's law results in reflection, four possible reflection type may occur according to given probability: Specular Spike, Specular Lobe, Lambertian and Backscattering (See Fig. \ref{fig:reflection}).

\begin{figure}[!htb]
\centering
\includegraphics[width=0.60\linewidth]{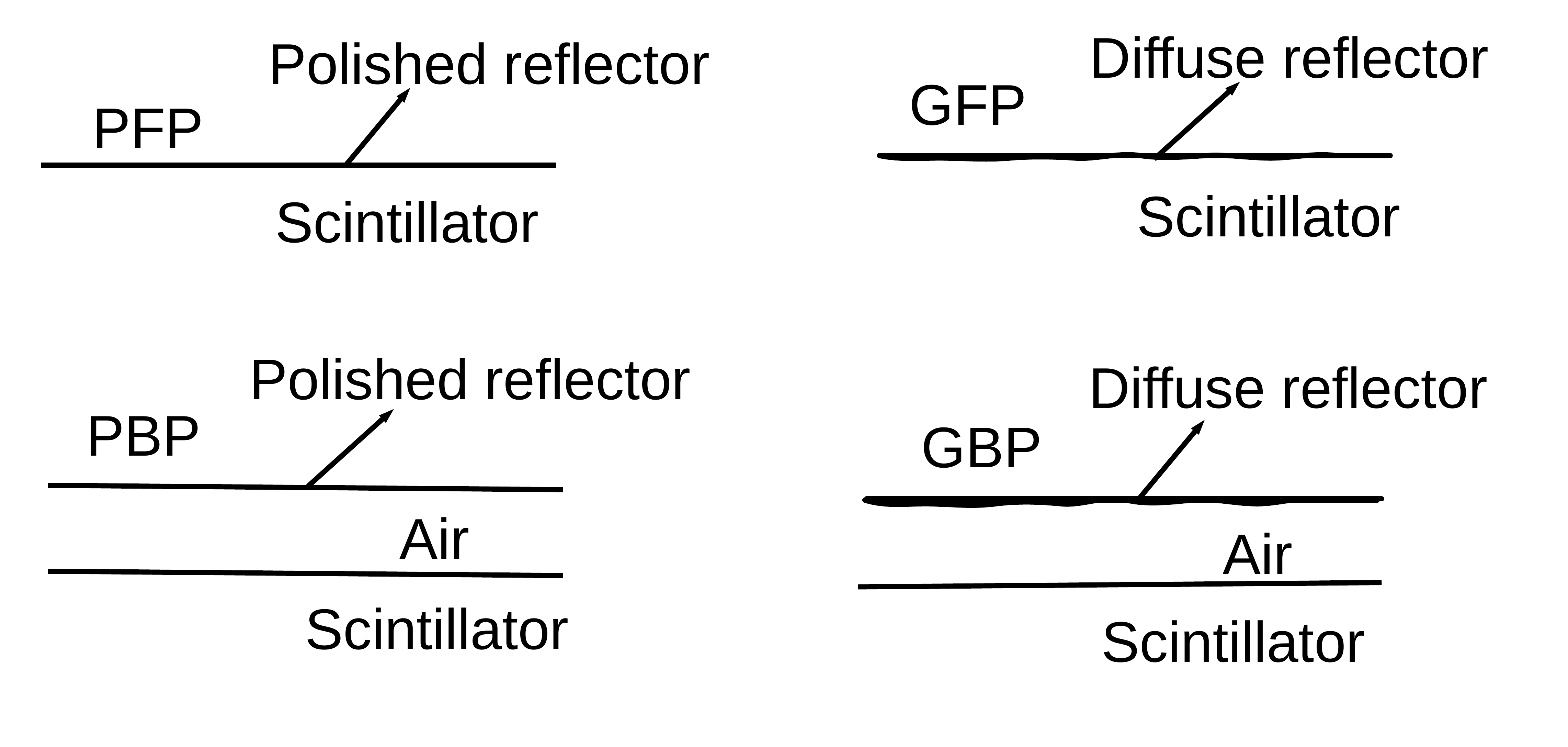}
\caption{ The parameter $\sigma_{\alpha}$ controls the degree of surface roughness between the scintillator and the air. If a photon reflects from the polished reflector, it is specular spike reflection. If a photon reflects from the diffuse reflector, it is Lambertian reflection. }
\label{fig:Wrap}
\end{figure}

With respect to reflector coating, the UNIFIED model offers two kinds of wrapping method for both polish and diffuse reflectors: if a reflector coated onto scintillator surface perfectly it is named front painted, on the other hand, if a reflector coated imperfectly onto scintillator surface by leaving an air gap between scintillator and reflector, it is named back painted. If we consider the reflector type, four parameters arise. These are PolishedFrontPainted (PFP), GroundFrontPainted (GFP), PolishedBackPainted (PBP), and GroundBackPainted (GBP). Visual representation of the parameters is shown in Fig. $\ref{fig:Wrap}$. These parameters will be used in our simulation to refer wrapping method for both polish and diffuse reflector. The detail explanation for each parameter is available in \cite{Agostinelli}.

The simulations are performed in the following orders: In section  $\ref{sec:geo}$, we investigate the effect of photocathode size and light guide shape on light collection efficiency (LCE) and detected spectrum. In section $\ref{sec:wrap}$, we determine the best reflector type and its applying method onto scintillator and light guides surface. In section $\ref{sec:ref}$, we compare the effect of reflectivity on LCE for some commonly used high reflectance material and finally in section $\ref{sec:sigma}$, we evaluate scintillator surfaces having different surface roughness degree.

\section{Geometry And Material }

\begin{figure}[!htb]
    \centering
    \subfigure[Module 1]
    {
        \includegraphics[width=0.50\linewidth]{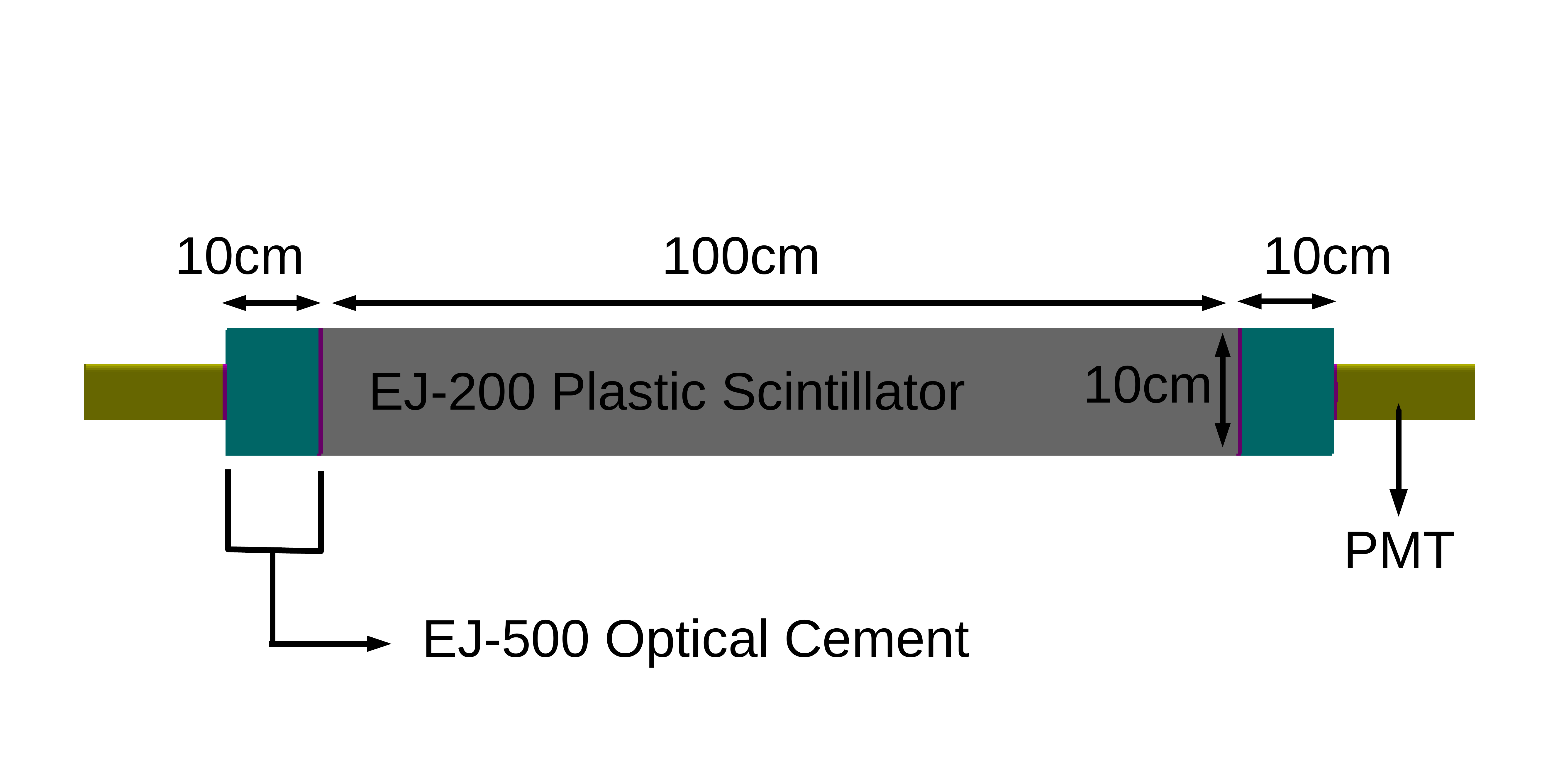}
        \label{fig:module1}
    }
    \subfigure[Module 2]
    {
       \includegraphics[width=0.47\linewidth]{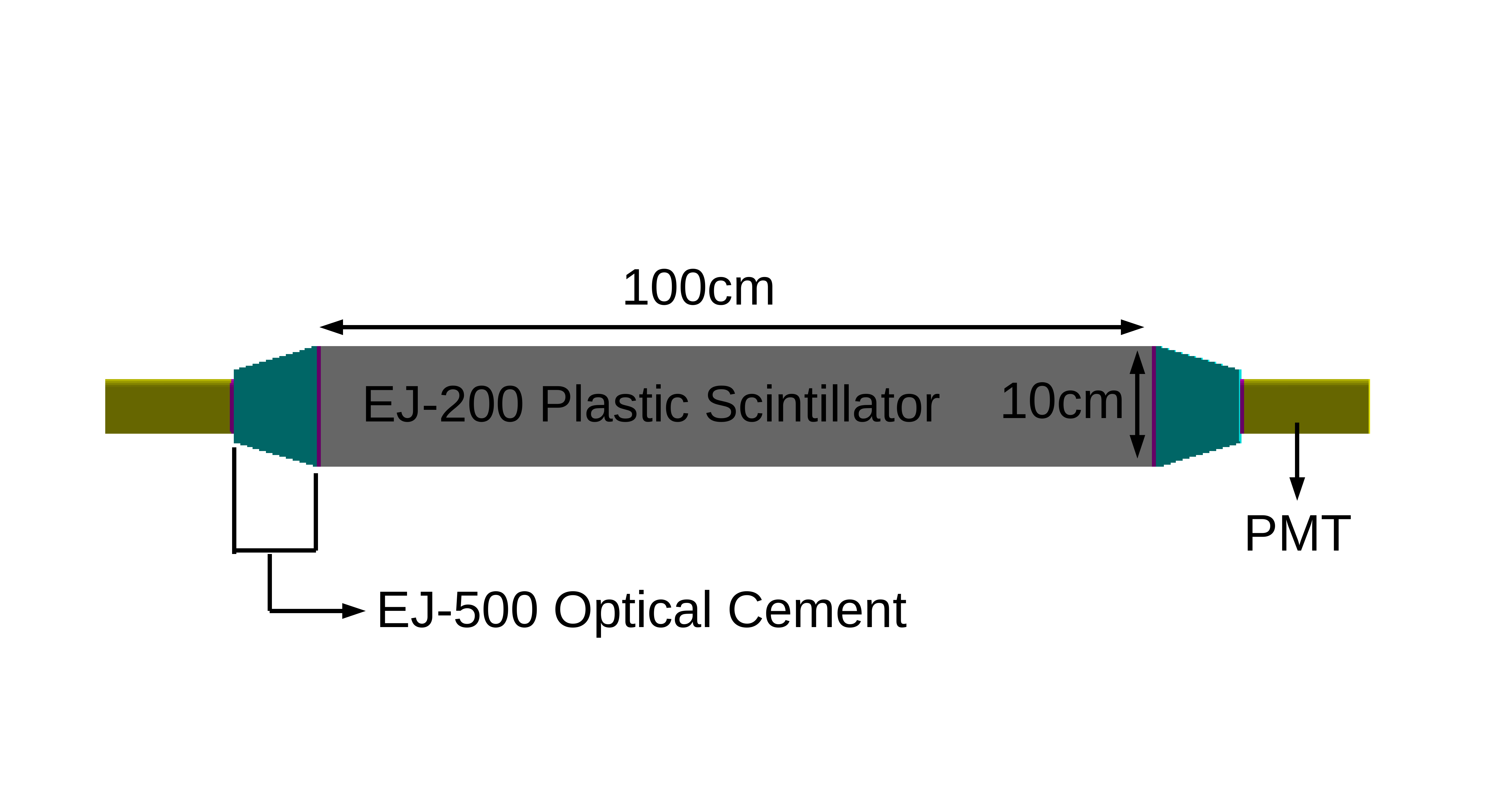}
        \label{fig:module2}
    }
    \subfigure[Module 3]
    {
       \includegraphics[width=0.47\linewidth]{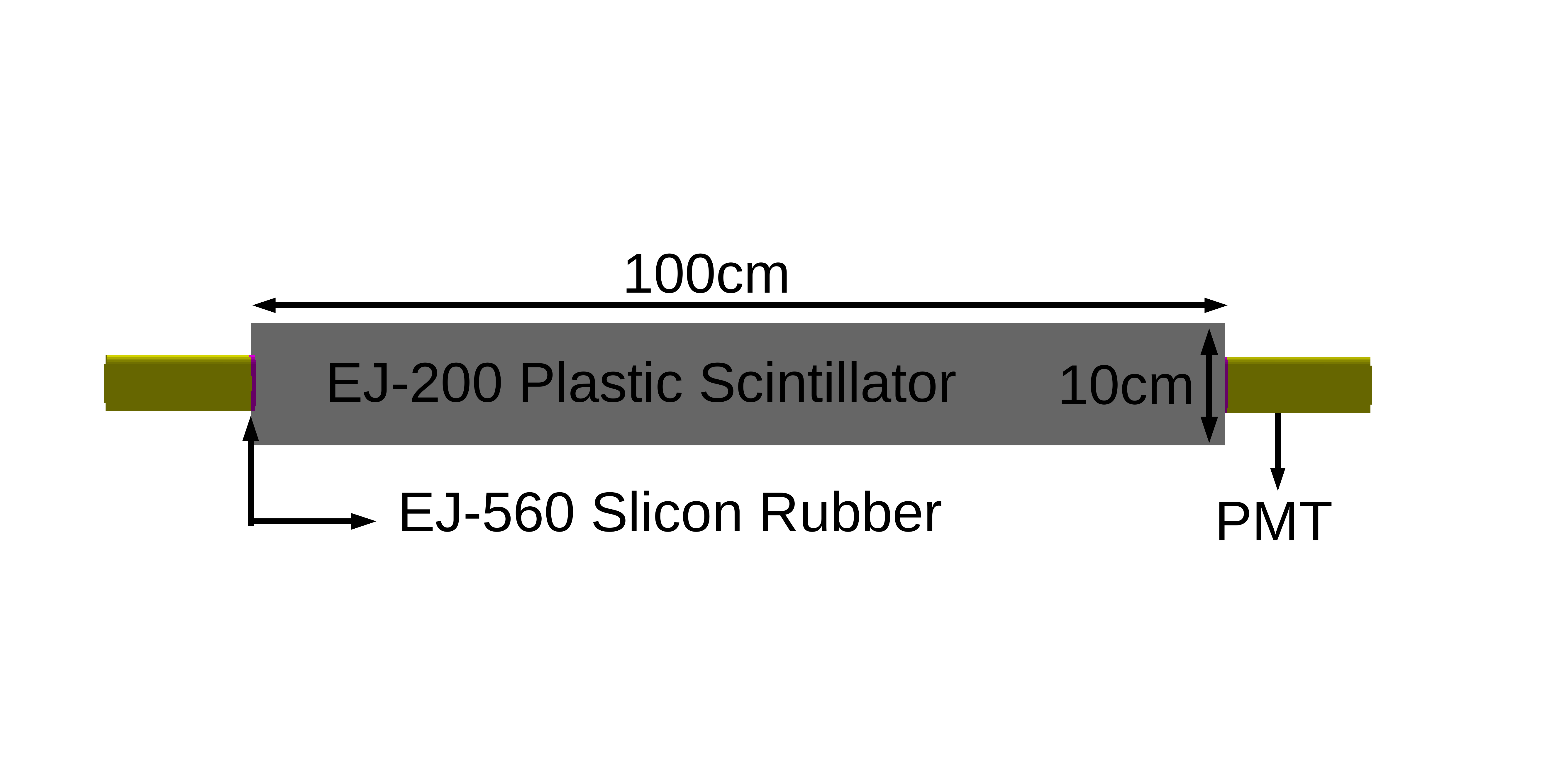}
        \label{fig:module3}
    }
    \caption{ Different type of antineutrino detector modules. These are modeled based on ref [7], [8]. An antineutrino detector is made of an array of these modules. Two different PMTs are used for each module: 2-inch H6410 Hamamatsu PMT and 3-inch 9265B ET Enterprise PMT. For module 2, the surface area of the light guide attached to PMT set to 6cmx6cm for 2-inch PMT and 8cmx8cm for the 3-inch PMT. }
    \label{fig:modules}

\end{figure}

We use three antineutrino detector module configurations based on the ref.  \cite{Oguri,Battaglieri} to perform the simulations. The first module consist of three parts: a 10cm$\times$10cm$\times$100cm plastic scintillator bar (EJ-200, ELJEN Technology \cite{Eljen}), two 10cm$\times$10cm $\times$10cm acrylic cubic light guides and two 2-inch photomultipliers (H6410, Hamamatsu \cite{Hamamatsu}) or 3-inch photomultipliers (9265B, ET Enterprises \cite{Et}). Light guides are glued to both ends of the plastic bar and then connected to photomultipliers (PMTs) from both sides with the help of optical cement (EJ-500). In the second module, we only modify light guide shape from cubic to trapezoid. Finally in the third module, we remove the light guide and directly couple the plastic scintillator bar to the photocathode of PMTs with silicone rubber (EJ-560). The schematic view of the modules is shown in Fig. \ref{fig:module1}, \ref{fig:module2} and \ref{fig:module3} respectively.

\begin{figure}[!htb]
\centering
\includegraphics[width=0.50\linewidth]{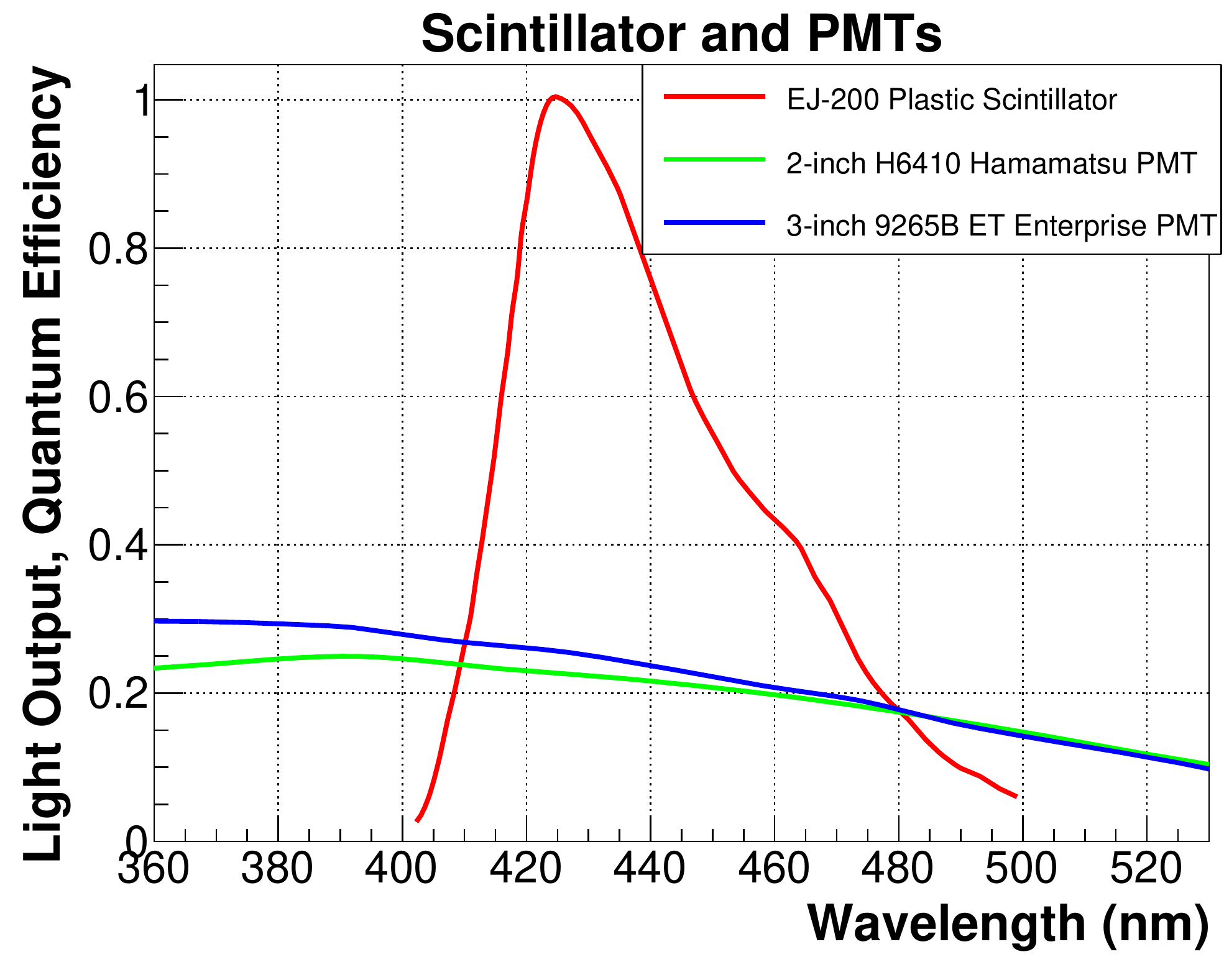}
\caption{ Emission spectrum of EJ-200 plastic scintillator and quantum efficiency of two different PMTs. }
\label{fig:ScinAndPMT}
\end{figure}

The material properties of the scintillator and important parameters used in the simulation are listed in table \ref{table:1}. The light emission spectrum of plastic scintillator and the quantum efficiency of PMT photocathode are shown in Fig. \ref{fig:ScinAndPMT}. These properties are added to the code to perform simulation properly. The red line shows the probability of light emission with a given wavelength when a particle deposit its energy. The blue and green lines show the detection probability of photons (quantum efficiency) according to the wavelength.  

\begin{table}[!htb]
    \center 
    \caption{Input parameters used in the simulation }
 \label{table:1}
\begin{tabular}{  p{9.2cm} p{2.1cm}  }

\hline
\quad \quad \quad  EJ-200 Material Properties   \\

Light output($\%Anthracene$)& 64  \\

Scintillation Yield ($\frac{photons}{MeV}$) &  10,000  \\

Wavelength of maximum emission ($nm$) &  425  \\

Light attenuation length ($cm$) & 380  \\

Density($\frac{g}{cm^3}$) & 1.023   \\

Refractive index & 1.58   \\

\hline
  
Refractive index of air & 1.00028 \\
Refractive index of light guide & 1.502 \\
Refractive index of EJ-500 optical cement & 1.57 \\
Refractive index of EJ-560 slicon rubber & 1.43 \\

\hline
\end{tabular}
    
\end{table}

\section{Methods}
\label{met}

Numerous parameters influencing the light collection of antineutrino detector modules are studied simulating the energy deposition of the positron, which is the product of inverse beta decay. Positron is generated randomly in a point inside the scintillator and fired at any direction with the energy of 1 MeV. Following the energy deposition of the positron in the scintillator, scintillation photons are generated and tracked throughout the detector until they disappear. (e.g., self-absorption in scintillator, losses at the scintillator surfaces, and detected by PMT) . The parameters affecting LCE are changed, and the effect of each parameter is examined by comparing the change in LCE value and the change in the detected spectrum, i.e. the spectrum of photoelectrons per positron depositing energy in the plastic scintillator. In addition, the effect of each parameter is also evaluated simulating the energy deposition of Cobalt-60 (1173keV and 1332 keV photon) and Cesium-137 (662KeV photon) gamma sources.

\begin{figure}[!htb]
\centering
\includegraphics[width=0.50\linewidth]{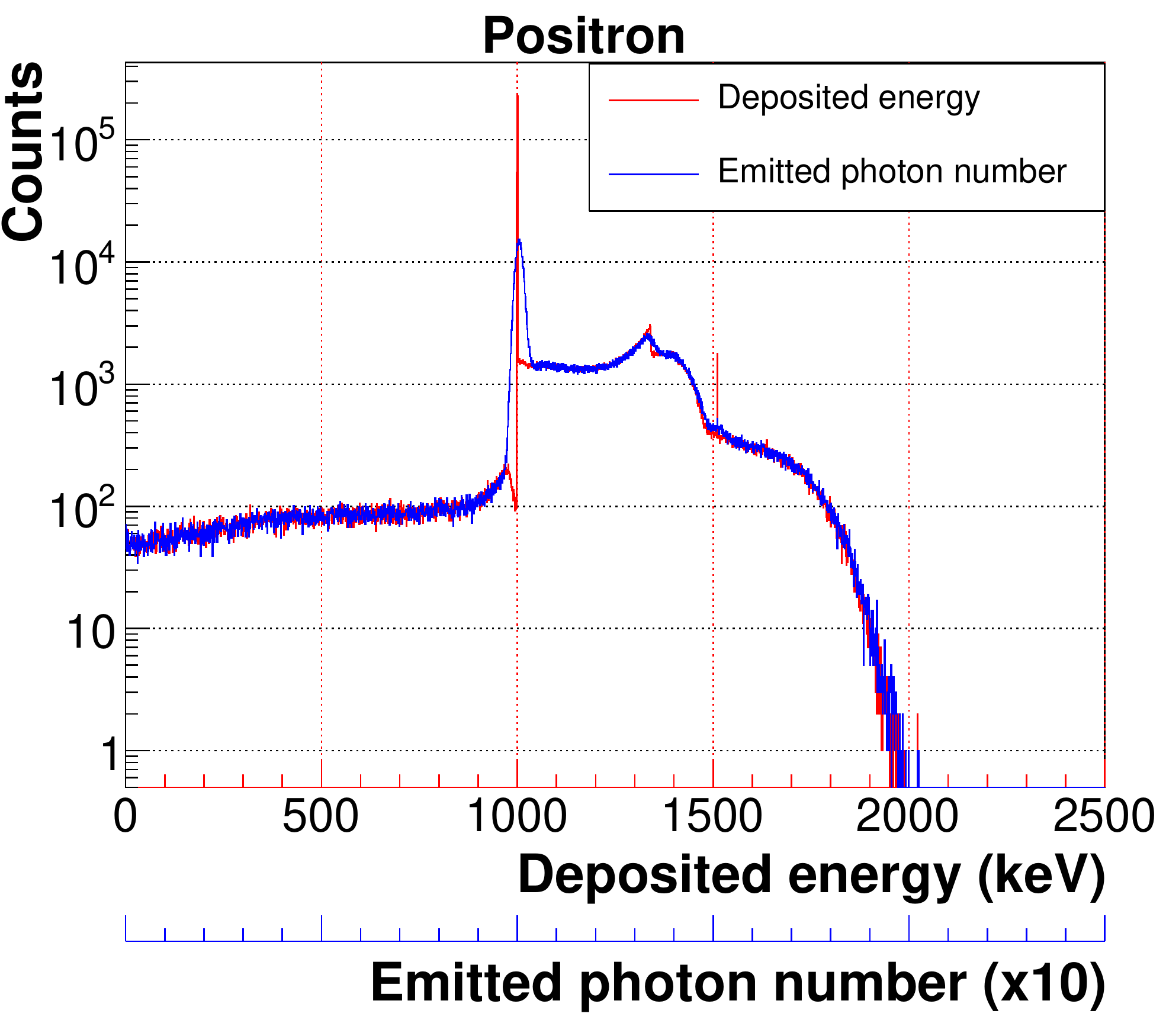}
\caption{ 
Simulated energy spectrum of 1 MeV positron.  When positron deposits E amount of energy in EJ-200 plastic scintillator, an average of $N = E(keV) \times (Scintillation yield = 10photons/KeV)$ photons are emitted with a standard deviation of $\sqrt{N}$. }
\label{fig:posde}
\end{figure}

Fig. $\ref{fig:posde}$ shows simulated energy deposition of 1 MeV positron (red line) and emitted photon number distribution (blue line) with respect to deposited energy. Positron deposits its energy via ionisation and when it is almost zero kinetic energy it runs into an electron and produces two 511 keV gamma photon in opposite directions. Due to the low atomic number and low density of scintillator, $33\%$ of all events both gamma escape from the scintillator without energy deposition. The remaining $67\%$ of events at least one of the two gamma deposit some part of its energy. The peak at energy 1341 keV (1MeV positron + Compton edge of 511 keV gammas) stems from the events in which one of the two gamma escapes from the module and the other scatters at 180 degrees from a free electron. On the other hand, the events that one of the two gammas deposits its total energy and the other escape without energy deposition (1MeV positron + 511 keV gammas) correspond to the peak value of 1511 keV. Finally, the event in which both gamma deposit its total energy occurs only twice in 1 million events.

\begin{figure}[!htb]
    \centering
    \subfigure[Simulated energy spectrum of 1173 and 1332 keV gammas. ]
    {
        \includegraphics[width=0.47\linewidth]{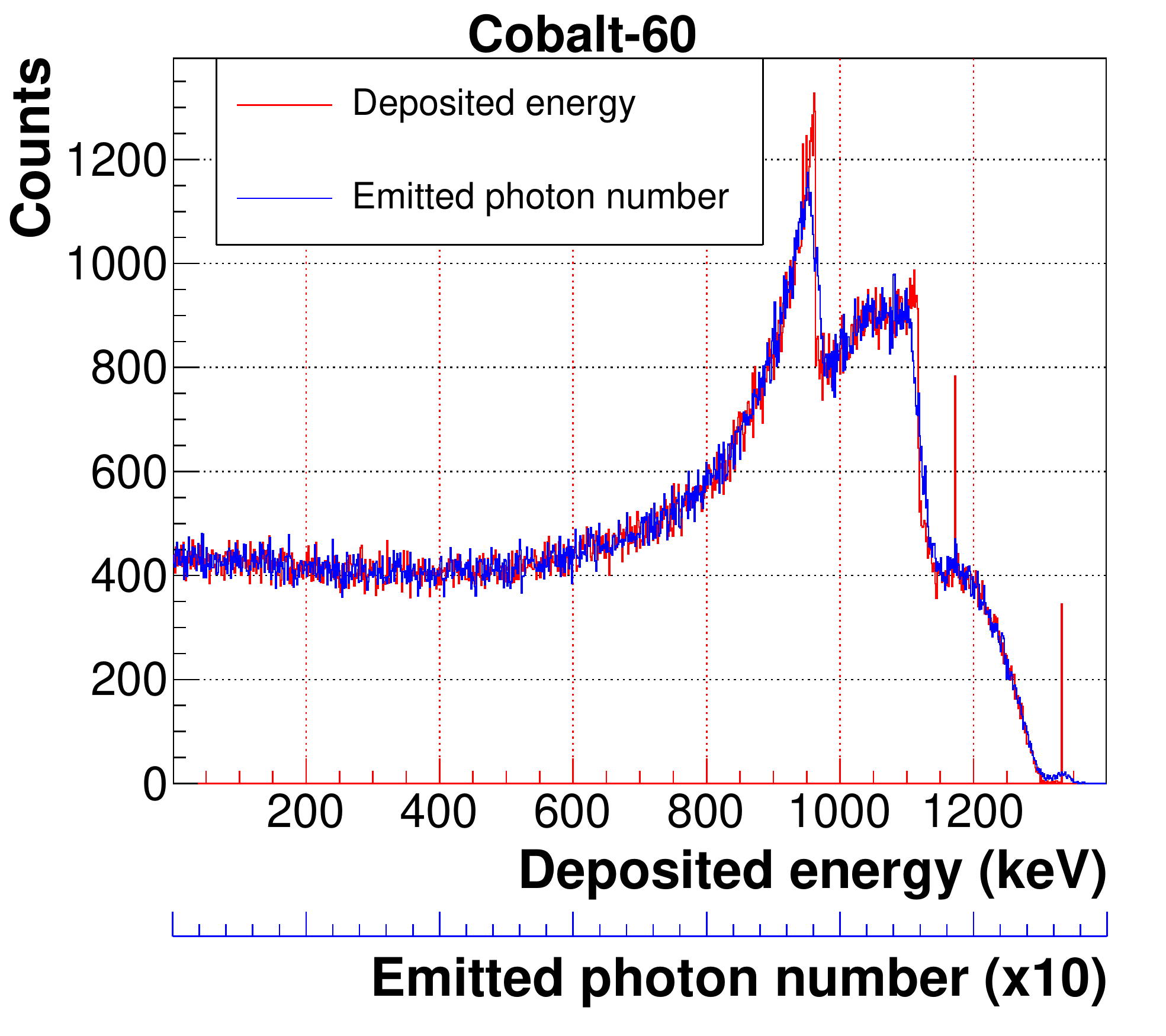}
        \label{fig:code}
    }
    \subfigure[Simulated energy spectrum of 662 keV gammas]
    {
       \includegraphics[width=0.47\linewidth]{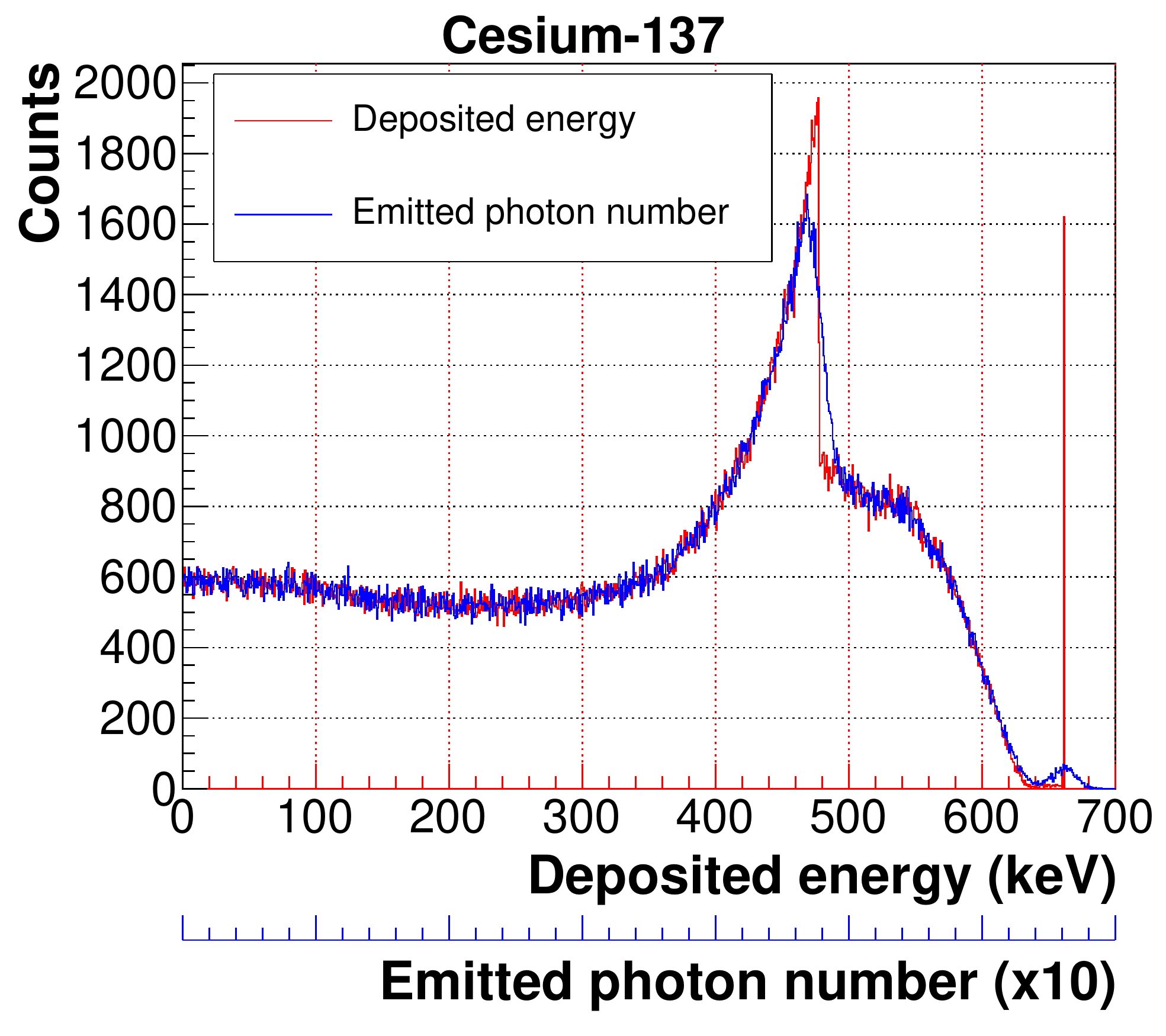}
        \label{fig:csde}
    }
    \caption{ Cobalt-60 and Cesium-137 gamma sources are located in the middle of the bottom surface of the detector. Following the energy deposition of gammas in the scintillator, optical photons are created and tracked throughout the detector module.  }
    \label{fig:cocs}
\end{figure}

    In order for our work to be experimentally comparable, the impact of each parameter on detected spectrum is also investigated simulating the energy deposition of $^{137}Cs$ and $^{60}Co$ gamma sources. Fig. $\ref{fig:cocs}$ shows simulated energy spectrum and associated emitted photon number distribution. These emitted scintillation photons are tracked in each simulation setup.

 % Figure 2 shows the emitted photon number distribution with respect to deposited energy(figure 1). The peak in figure1 is related to compton events and full energy deposition.   

\section{Results}
\subsection{Impact of light guide shape and photocathode size on LCE and detected spectrum}  
\label{sec:geo}

\begin{figure}[!htb]
\centering
\includegraphics[width=0.50\linewidth]{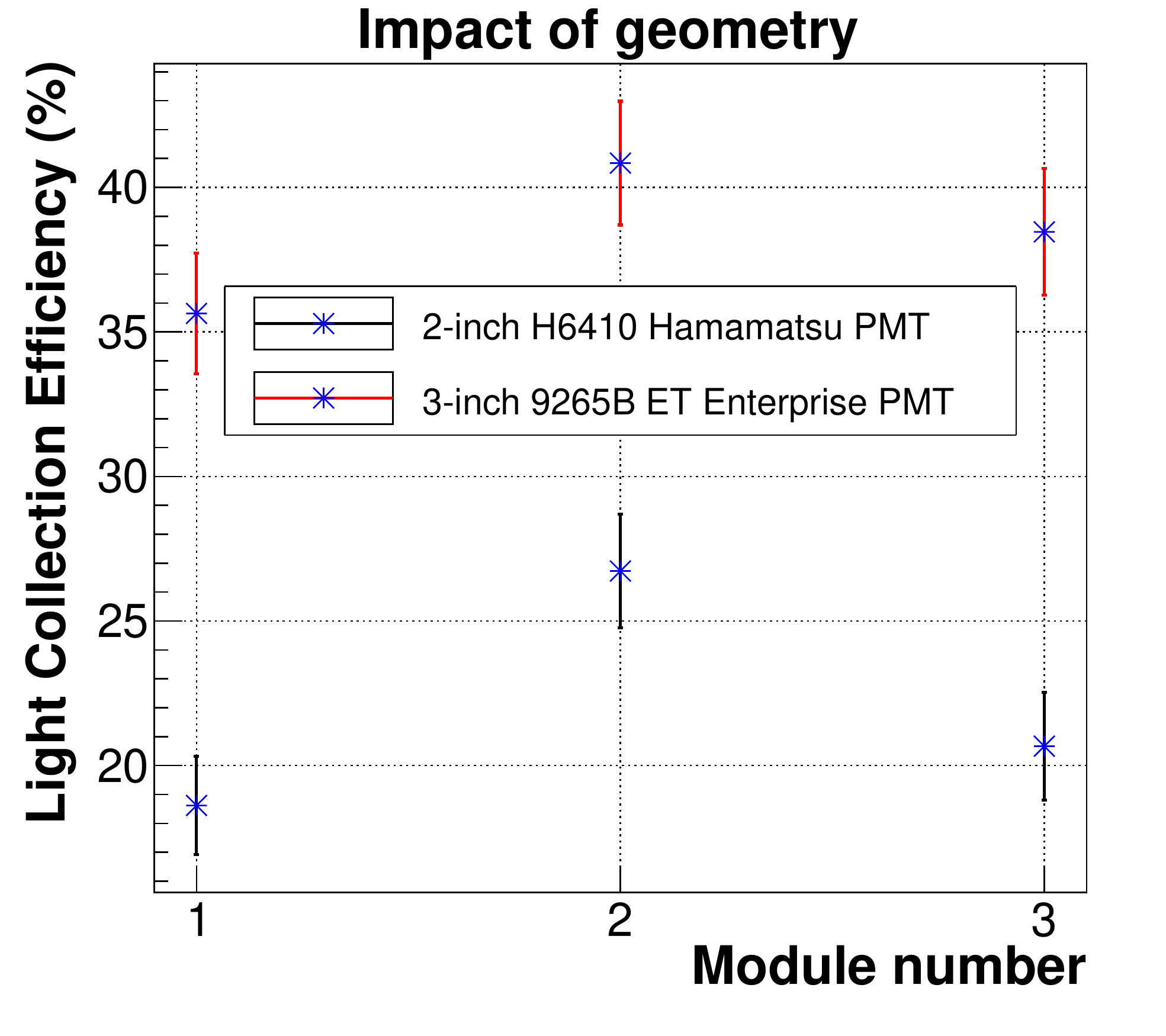}
\caption{ The impact of light guide shape and photocathode size on light collection efficiency. The effective photocathode size (radius) for 2-inch H6410 Hamamatsu PMT \cite{Hamamatsu} and 3-inch 9265B ET Enterprise PMT \cite{Et} are set to 2.3 and 3.5 respectively. The scintillator surface finish type set to PBP and the parameter indicating the degree of scintillator surface polishing level ($\sigma_{\alpha}$) set to 0.2. The reflectivity of reflector (R) set to 0.98. }
\label{fig:MODULES}
\end{figure}

To investigate the effect of light guide shape and photocathode size on light collection efficiency (ratio of the number of photons arrived at photocathode surface to the number of photons emitted), we consider three types of antineutrino detector module: module 1, module 2 and module 3 (See Fig. $\ref{fig:modules}$). The modules differ from each other only in terms of light guide shape. The simulation is performed using two different PMTs for each module: 2-inch H6410 Hamamatsu PMT and 3-inch 9265B ET Enterprise PMT. The results are shown in Fig. $\ref{fig:MODULES}$. In the figure, we can see that the highest light collection efficiency is obtained when module 2 is used. Because adjusting the surface area of the light guide according to PMT surface increases the LCE and decreases the variation in LCE. The effect of light guide shape on LCE is more pronounced when 2-inch PMT is used. In this case, module 2 collects 8$\%$ more light than module 1 and 6$\%$ more light than module 3. On the other hand, in the case of using 3-inch PMT module 2 collects 5$\%$ more light than module 1 and 3$\%$ more light than module 3. In addition, a spectacular increase in LCE arises from the size of the photocathode. When the radius of the photocathode is increased from 2.3cm (2-inch H6410) to 3.5cm (3-inch 9265B), the LCE increases from 19$\%$ to 36$\%$, 27$\%$ to 41$\%$ and 21$\%$ to 37$\%$ for module 1, module 2 and module 3 respectively. 

\begin{figure}[!htb]
    \centering
    \subfigure[Positron]
    {
        \includegraphics[width=0.50\linewidth]{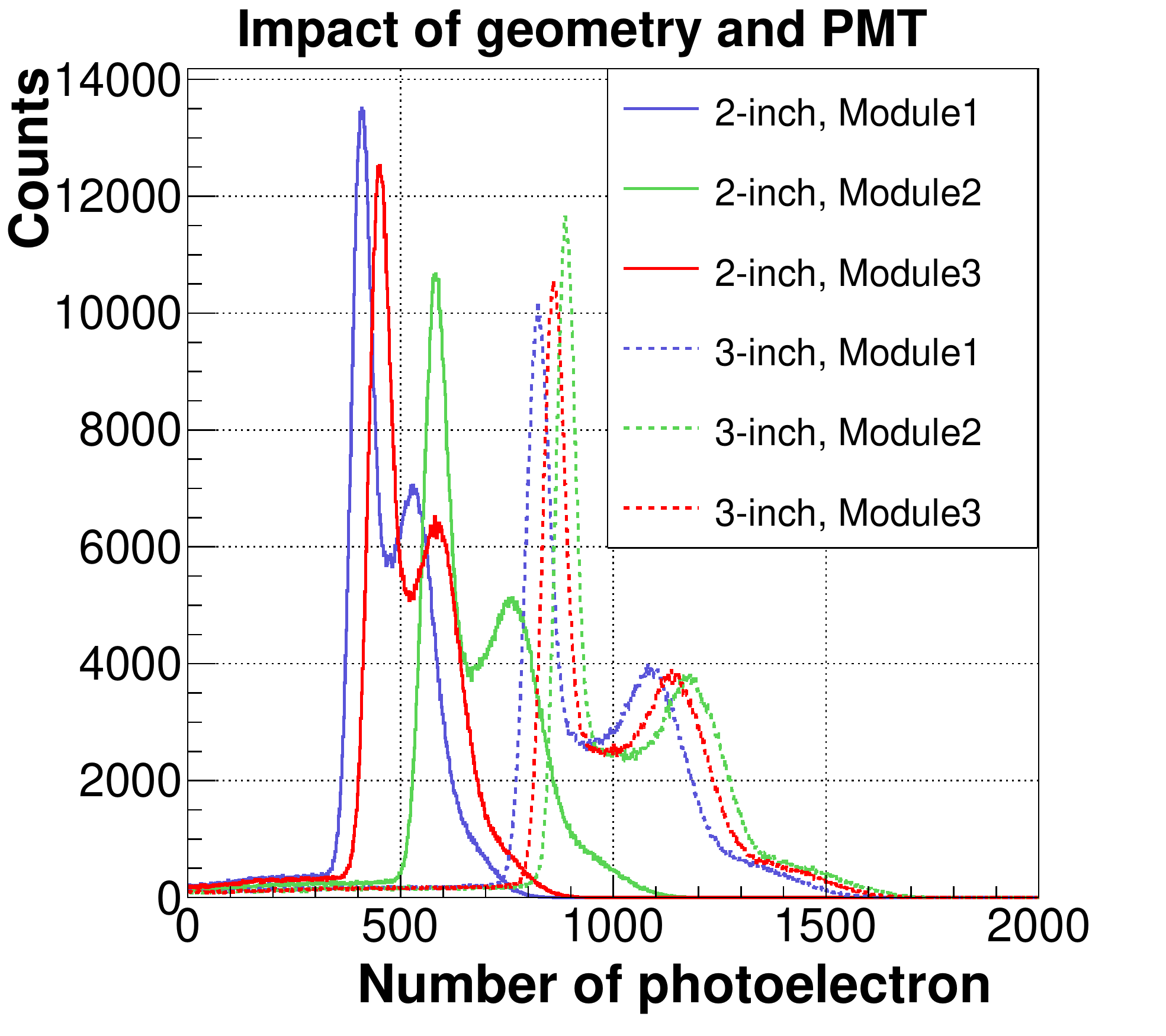}
        \label{fig:posmod}
    }
    \\
     \subfigure[$^{60}Co$]
    {
       \includegraphics[width=0.47\linewidth]{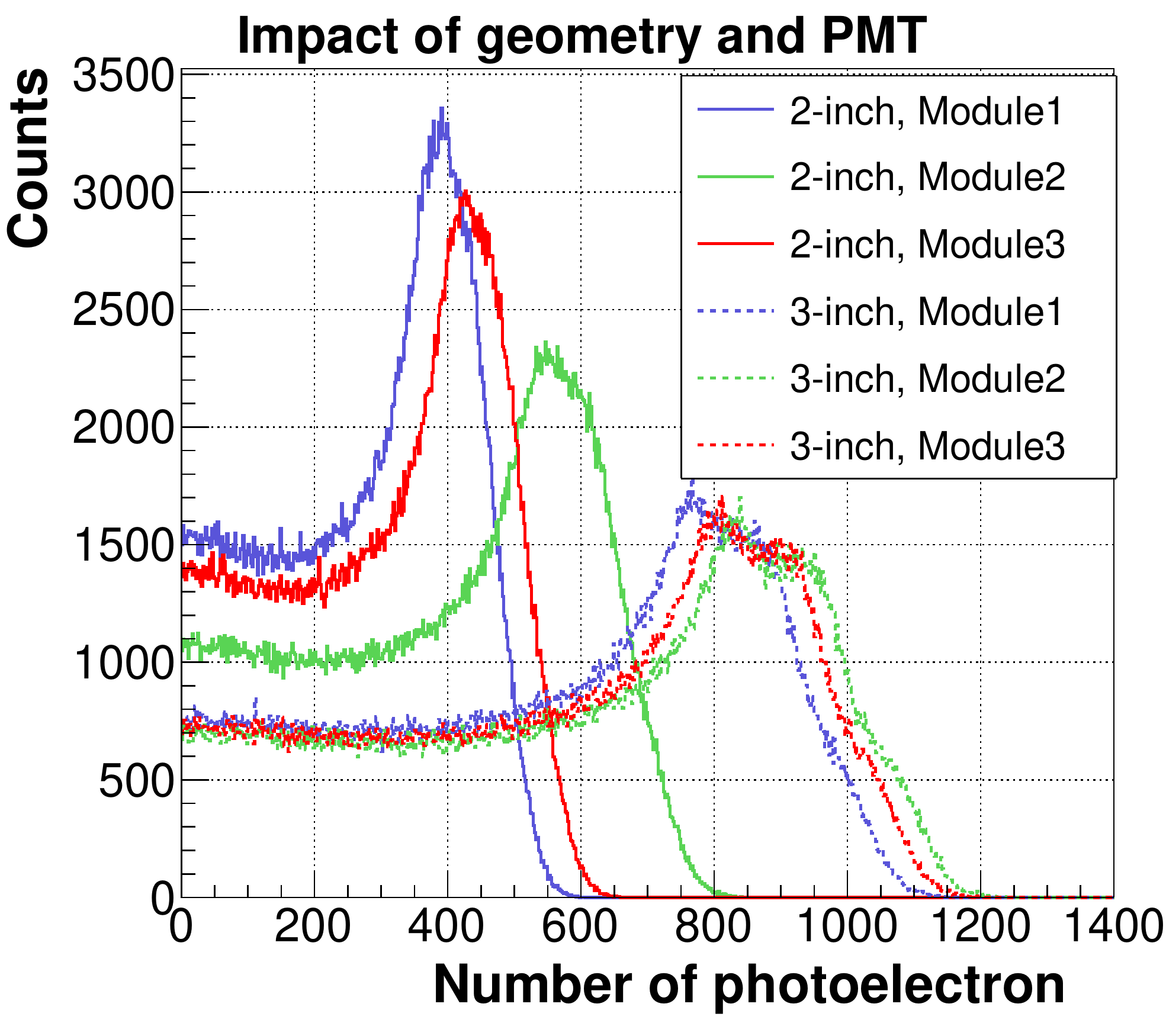}
        \label{fig:comod}
    }
    \subfigure[$^{137}Cs$]
    {
       \includegraphics[width=0.47\linewidth]{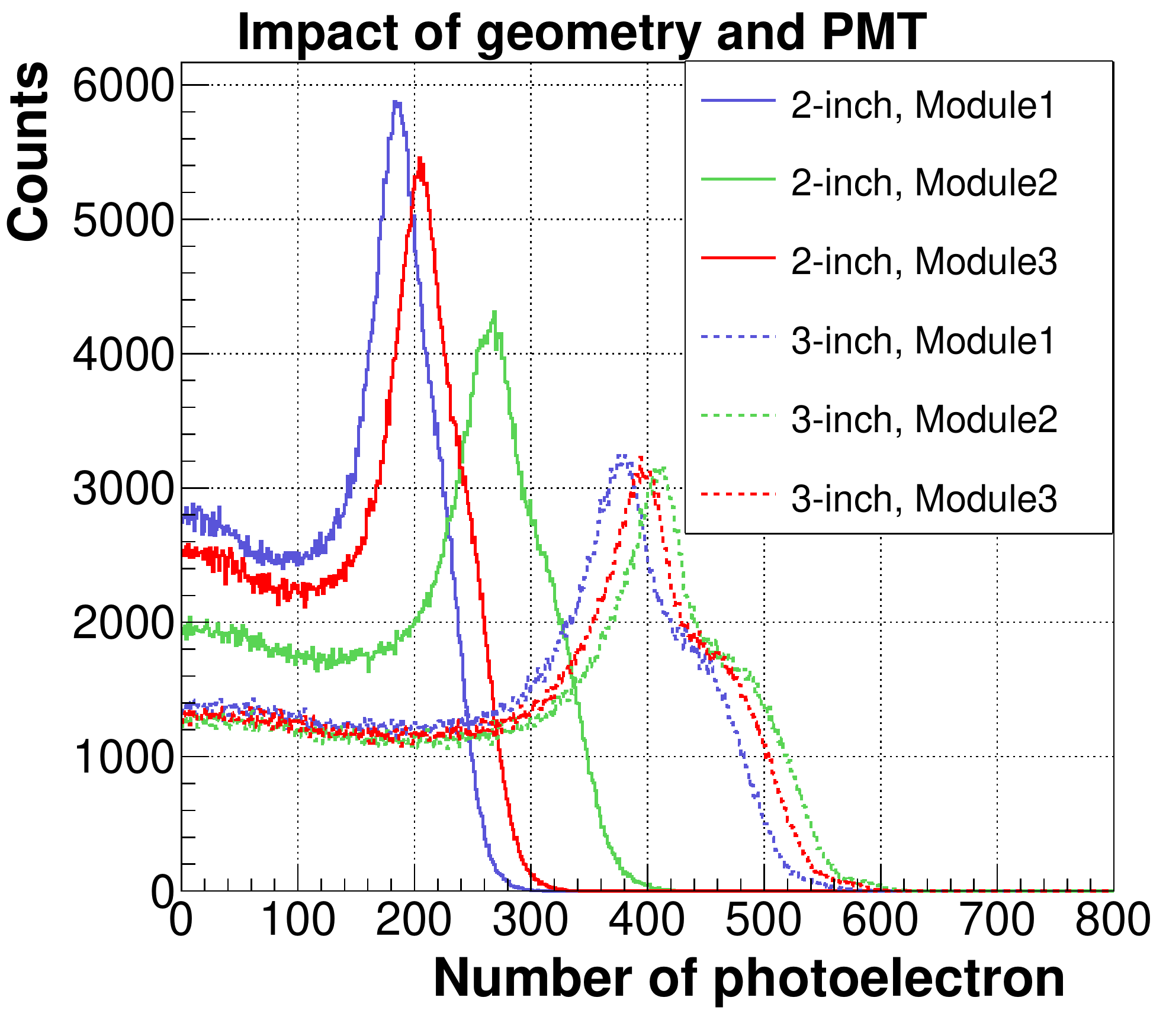}
        \label{fig:csmod}
    }
    \caption{ The comparison of photoelectron spectra for three modules in the case of using 2-inch H6410 Hamamatsu PMT and 3-inch 9265B ET Enterprise PMT. The modules are wrapped slightly with the specular reflector. The reflectivity of reflector (R) set to 0.98 and the parameter $\sigma_{\alpha}$ set to 0.2.}
    \label{fig:MOD}
      
\end{figure}

Fig. $\ref{fig:MOD}$ compares the photoelectron spectrum of each module in the case of using two different PMTs: 2-inch H6410 Hamamatsu PMT and 3-inch 9265B ET Enterprise PMT. The simulation is conducted simulating the energy deposition of 1 MeV positron (Fig. $\ref{fig:posmod}$), $^{60}Co$ (Fig. $\ref{fig:comod}$) and $^{137}Cs$ (Fig. $\ref{fig:csmod}$). For the case of using 2-inch H6410 Hamamatsu PMT in Fig. $\ref{fig:posmod}$, with respect to module 1, the peak position of module 2 and module 3 shifts to 43$\%$ and 10$\%$ to the right respectively. On the other hand, for the case of using 3-inch 9265B ET Enterprise PMT, with respect to module 1, the peak position of module 2 and module 3 shifts to 8$\%$ and 5$\%$ to the right respectively.

\subsection{Determination of best efficient reflector type and its applying method onto scintillator and light guide surface.}
\label{sec:wrap}

In order to determine the best reflector type and its applying method onto scintillator and light guide surface, we consider the cases where module 1 is covered loosely (BP) and tightly (FP) with both specular (PFP, PBP) and diffuse reflector (GFP, GBP). For each case, we use positron (Fig. $\ref{fig:posde}$), $^{60}Co$ and $^{137}Cs$ emitted photon number distributions (Fig. $\ref{fig:cocs}$).
 
\begin{figure}[!htb]
    \centering
    \subfigure[Positron]
    {
        \includegraphics[width=0.50\linewidth]{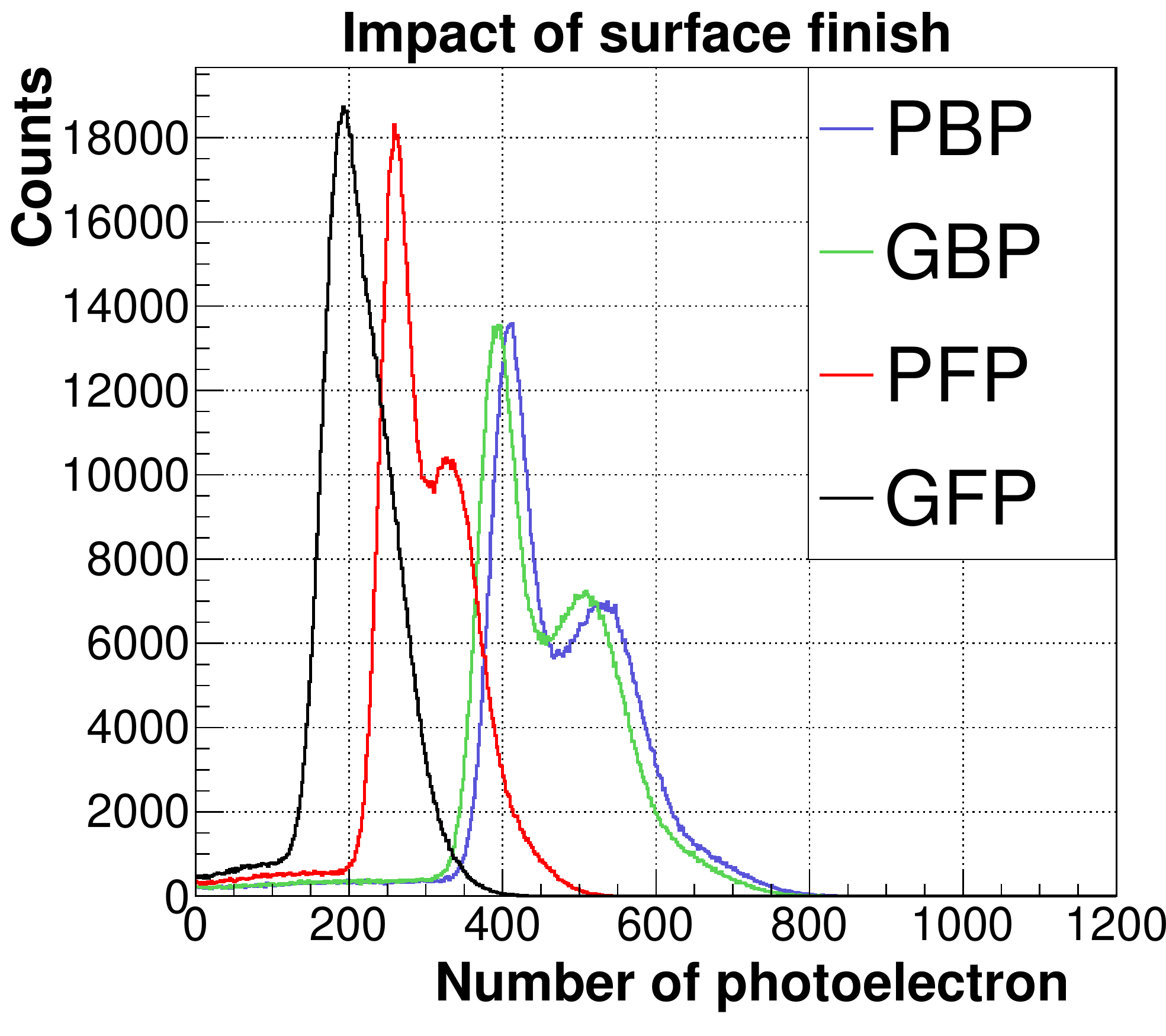}
        \label{fig:possf}
    }
    \\
    \subfigure[$^{60}Co$]
    {
       \includegraphics[width=0.47\linewidth]{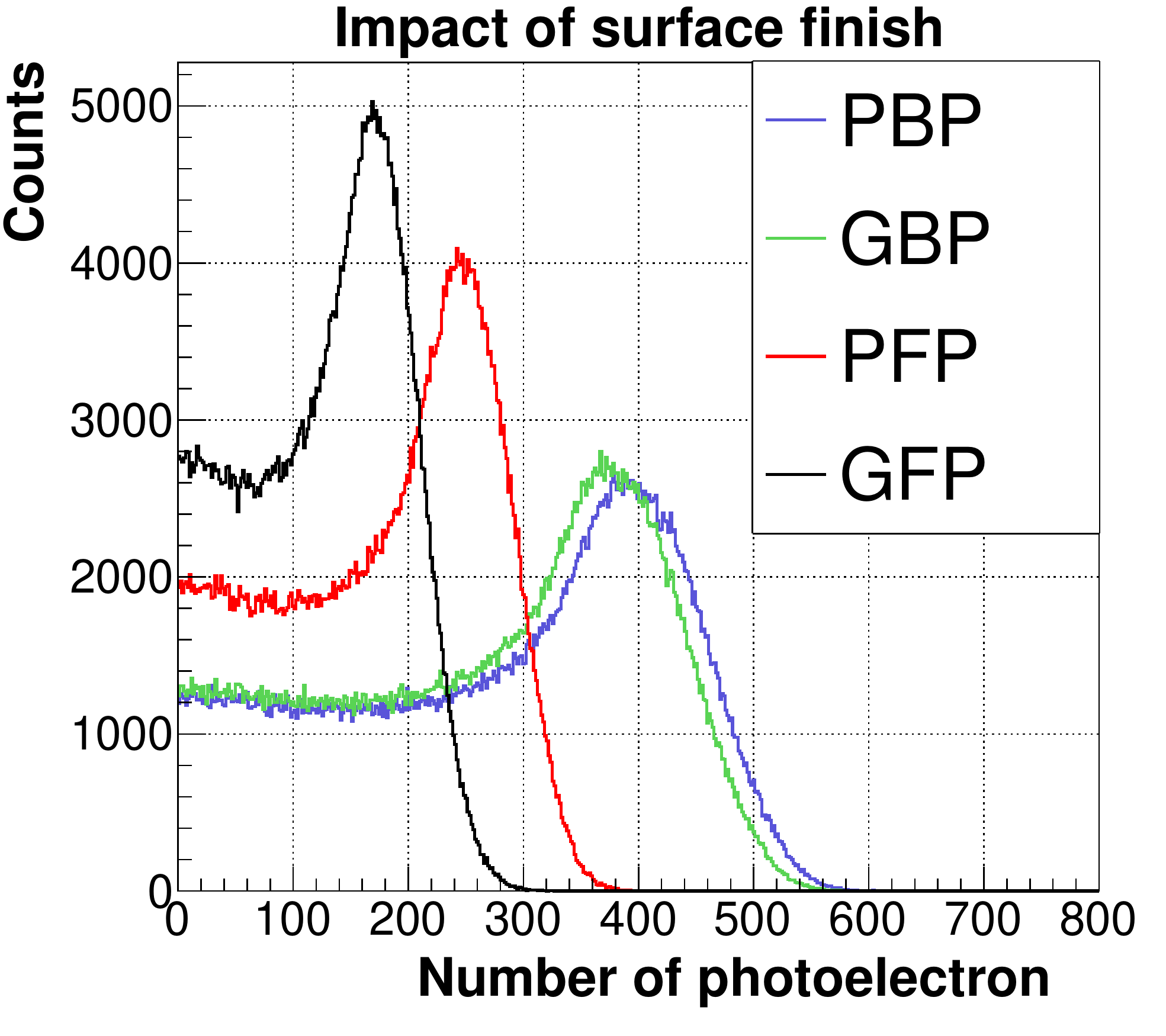}
        \label{fig:cosf}
    }
    \subfigure[$^{137}Cs$]
    {
       \includegraphics[width=0.47\linewidth]{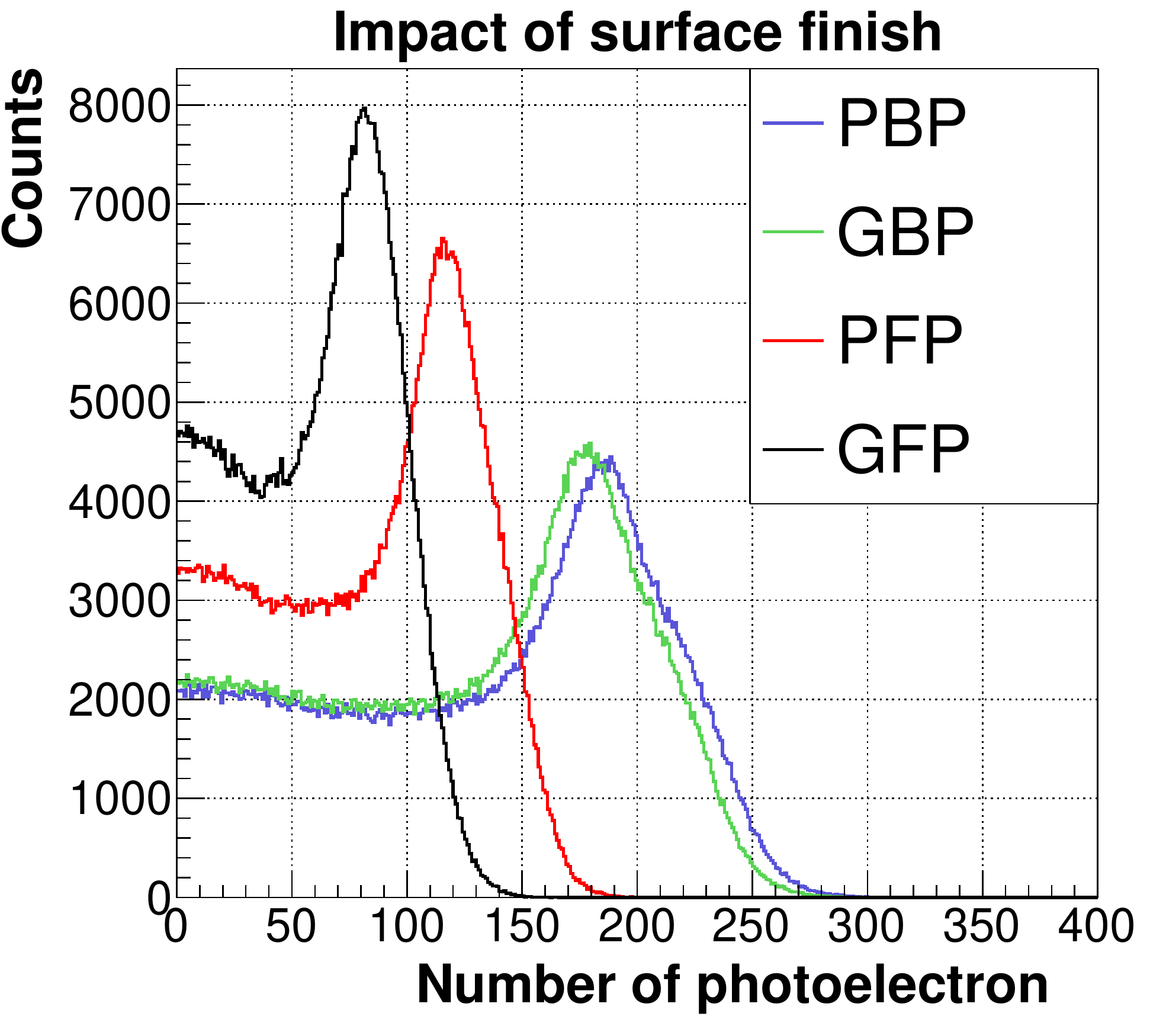}
        \label{fig:cssf}
    }
    \caption{ The impact of reflector type and its applying method onto scintillator and light guide surface. PFP and GFP represent the cases where module 1 is coated perfectly with specular and diffuse reflector respectively without any gap.  PBP and GBP represent the cases where module 1 is wrapped simply with specular and diffuse reflector respectively. In this case, there is an air gap between scintillator-light guide and reflector. 2-inch H6410 Hamamatsu PMT is used in the simulation. As an input parameter, we set the reflectivity of reflector (R) to 0.98 and the parameter $\sigma_{\alpha}=0.2$ which indicate scintillator surface roughness degree. }
    \label{fig:SF}
    
\end{figure}

Fig. $\ref{fig:SF}$ shows the photoelectron spectra for 1 MeV positron ( Fig. $\ref{fig:possf}$), $^{60}Co$ ( Fig. $\ref{fig:cosf}$) and $^{137}Cs$ ( Fig. $\ref{fig:cssf}$) sources. All three figures clearly show that back painted method exhibits more performance than front painted method. With respect to PFP, the peak position of spectra (PBP) shifts by $58\%$ to the right. Because leaving an air gap between scintillator and reflector increases the total internal reflection probability and thereby the number of photons arrived at photocathode surface. With regard to reflector type, specular reflector collects more light than diffuse reflector for both front painted and back painted method. For the back painted method, the peak position of spectra (PBP) shifts by $4\%$ to the right with respect to GBP. However, the difference are more pronounced when front painted method is used. Using a specular reflector (PFP) instead of a diffuse reflector (GFP) shifts the peak position of spectra by $40\%$ to the right.

\begin{figure}[!htb]
    \centering
    \subfigure[Simulated gamma spectra of different surface finish type for $^{60}Co$ (1173keV and 1332 keV photon) source. ]
    {
       \includegraphics[width=0.45\linewidth]{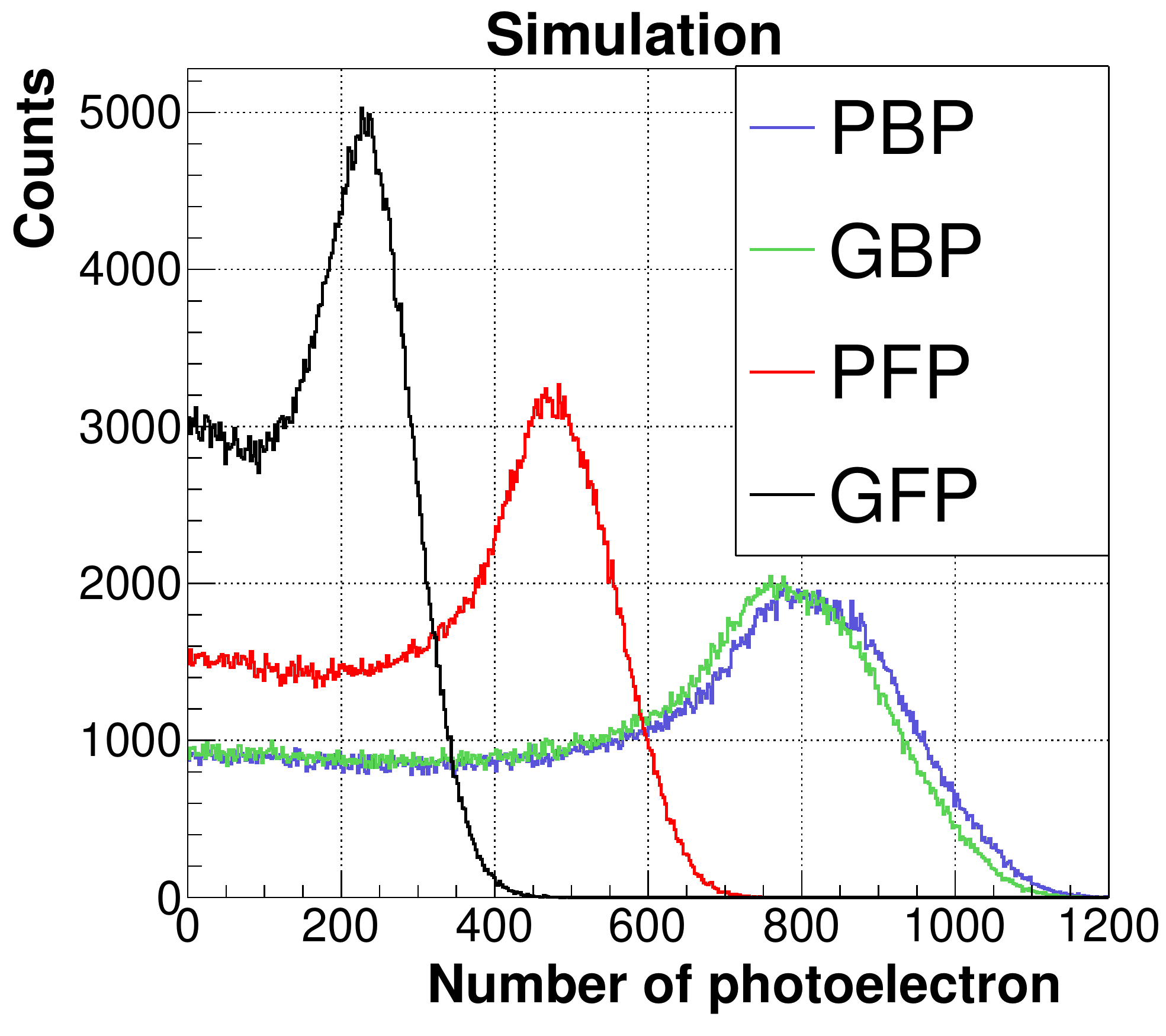}
        \label{fig:sim}
    }
    \quad
    \subfigure[The gamma spectra of different reflectors for $^{60}Co$ source at 20cm distance from the PMT \cite{taheri}.  ]
    {
        \includegraphics[width=0.45\linewidth]{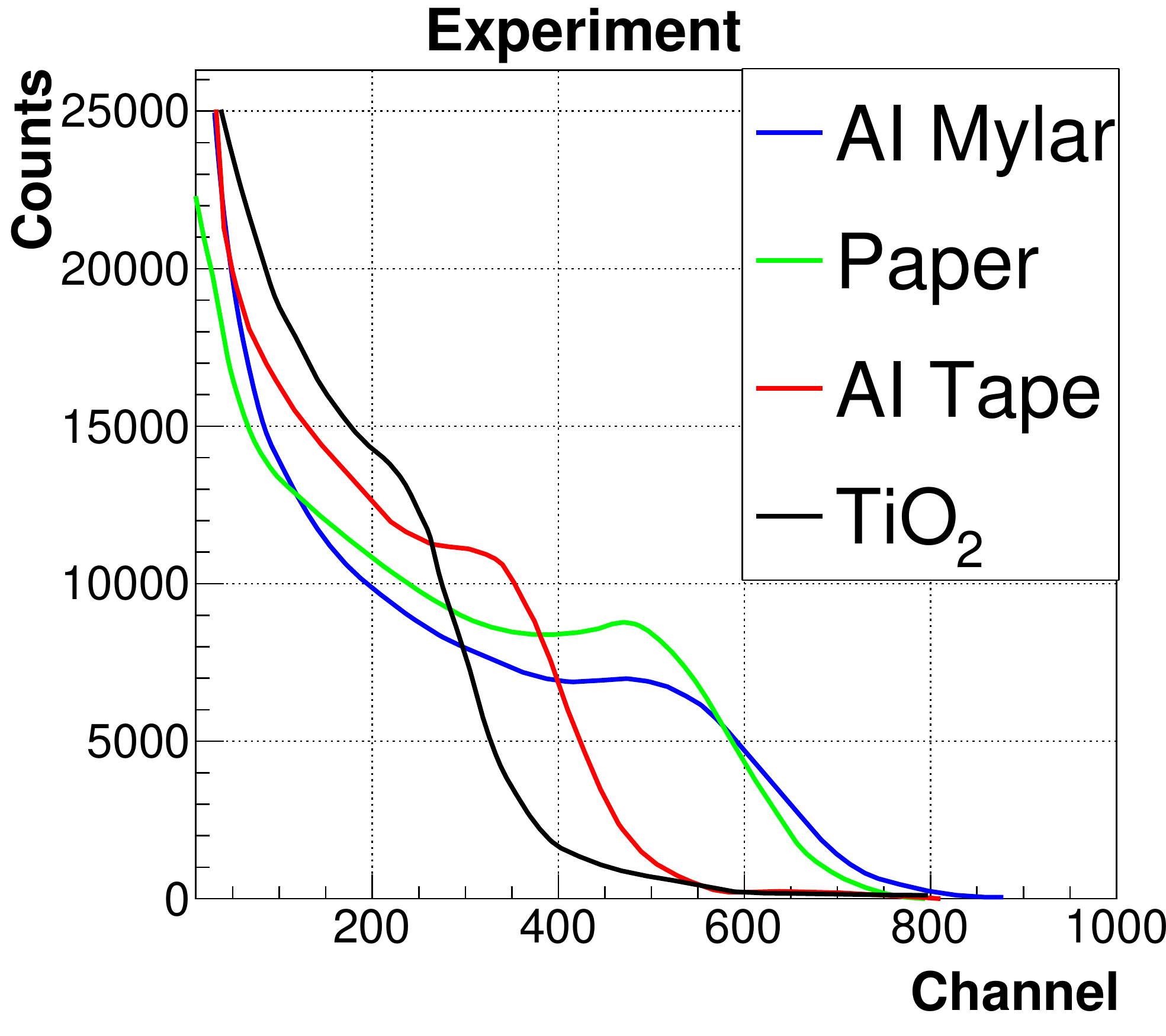}
        \label{fig:exp}
    }
     \caption{A comparison between the simulation and experiment. The peak position of the spectra determines the ability of the light collection of the reflector.}
    \label{fig:comp}
    
\end{figure}

To evaluate the validity of the simulation, it is compared with an experiment in which a rod shaped plastic scintillator with the diameter of 5.08 cm and length 50 cm is covered with 4 different reflective materials \cite{taheri}: TiO$_2$, AI-foil tape, AI Mylar and paper. In the experiment, these reflectors are applied to the scintillator surface using two different wrapping methods. TiO$_2$ and AI-foil tape are attached to scintillator surface firmly (which correspond to PFP and GFP parameters in our simulation) while AI Mylar and paper are loosely wrapped around the scintillator (which correspond to PBP and GBP parameters in our simulation). Cobalt-60 gamma source is used to generate optical photons in the scintillator. The simulation parameters are adjusted according to the experiment.\footnote{With regard to $^{60}Co$ gamma source positions, four different distances from the PMT is considered in the experiment: 10, 20, 30, 40 cm. But we only deal with a distance of 20 cm from the PMT for comparison with our simulation.} Fig. $\ref{fig:comp}$ compares the results of the experiment with our simulation. It is clear that the highest light collection is obtained respectively with AI Mylar (PBP), white paper (GBP), AI-foil tape (PFP) and TiO$_2$ (GFP), which is consistent with our simulation results.

\subsection{Impact of reflectivity on LCE and detected spectrum}
\label{sec:ref}
\begin{figure}[!htb]
\centering
\begin{minipage}[t]{.47\textwidth}
  \centering
  \includegraphics[width=1.0\linewidth]{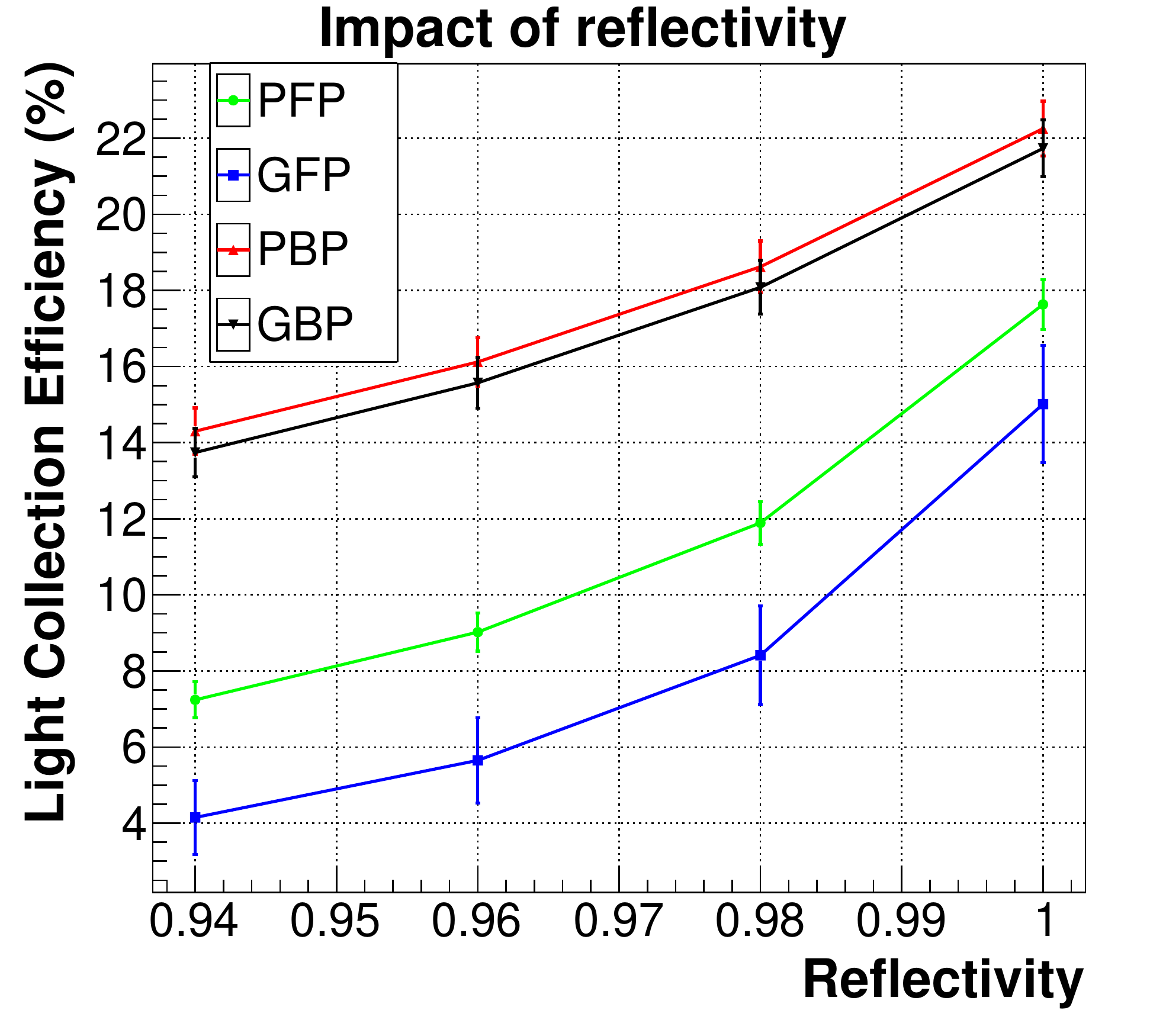}
 \caption{ The impact of reflectivity on light collection efficiency (LCE) in the case of four different surface finish. Here, the parameter R indicates the probability of a photon reflection from the reflecting material. If the photon is not reflected, it is absorbed. Module 1 and 2-inch H6410 Hamamatsu PMT photocathode size (diameter 4.6cm) is used to perform simulation. }
\label{fig:LCEREF}
\end{minipage}%
\quad 
\begin{minipage}[t]{.47\textwidth}
  \centering
  \includegraphics[width=1.0\linewidth]{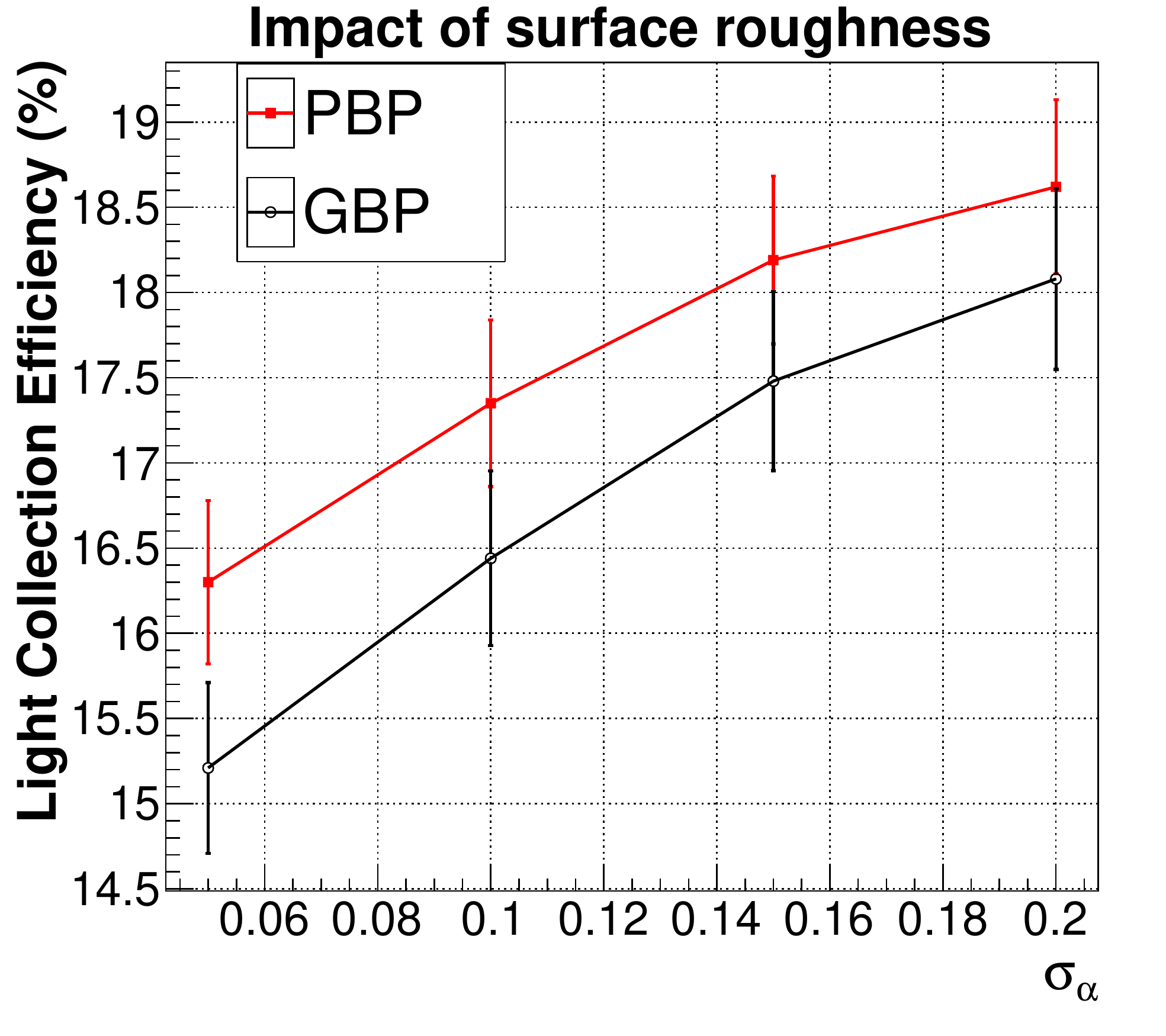}
  \caption{ The impact of scintillator surface roughness on light collection efficiency in the case of two different surface finish. The parameter $\sigma_{\alpha}$ indicates the degrees of surface roughness. Increasing the surface roughness more than 0.2 gives us less LCE value. The reflectivity of reflector (R) set to 0.98. }
\label{fig:LCESA}
\end{minipage}
\end{figure}

The effect of reflectivity on light collection efficiency is investigated in the case of four different surface finish: PFP, GFP, PBP and GBP (see Fig.$\ref{fig:Wrap}$). To perform simulation, module 1 is selected and the reflectivity value (R) of reflector set to 0.94, 0.96, 0.98 and 1.0. \footnote{ we consider some common reflector material reflectivity value from the reference \cite{Janecek}} For each reflectivity value, average light collection efficiency and its standard deviation are calculated. The results are shown in Fig. $\ref{fig:LCEREF}$. From the figure, it is obvious that LCE is significantly affected by the reflectivity value of reflecting material especially at higher reflectivity value. For example, when the back painted method is used, changing the reflectivity value of the reflector from 0.94 to 1.0 by equal 0.02 intervals increases the light collection efficiency value by about $13\%$, $16\%$ and $19\%$ respectively. Another important point is that the highest LCE is achieved for each reflectivity value when module 1 is wrapped simply with the specular reflector (PBP).  

\begin{figure}[!htb]
    \centering
    \subfigure[Positron]
    {
        \includegraphics[width=0.50\linewidth]{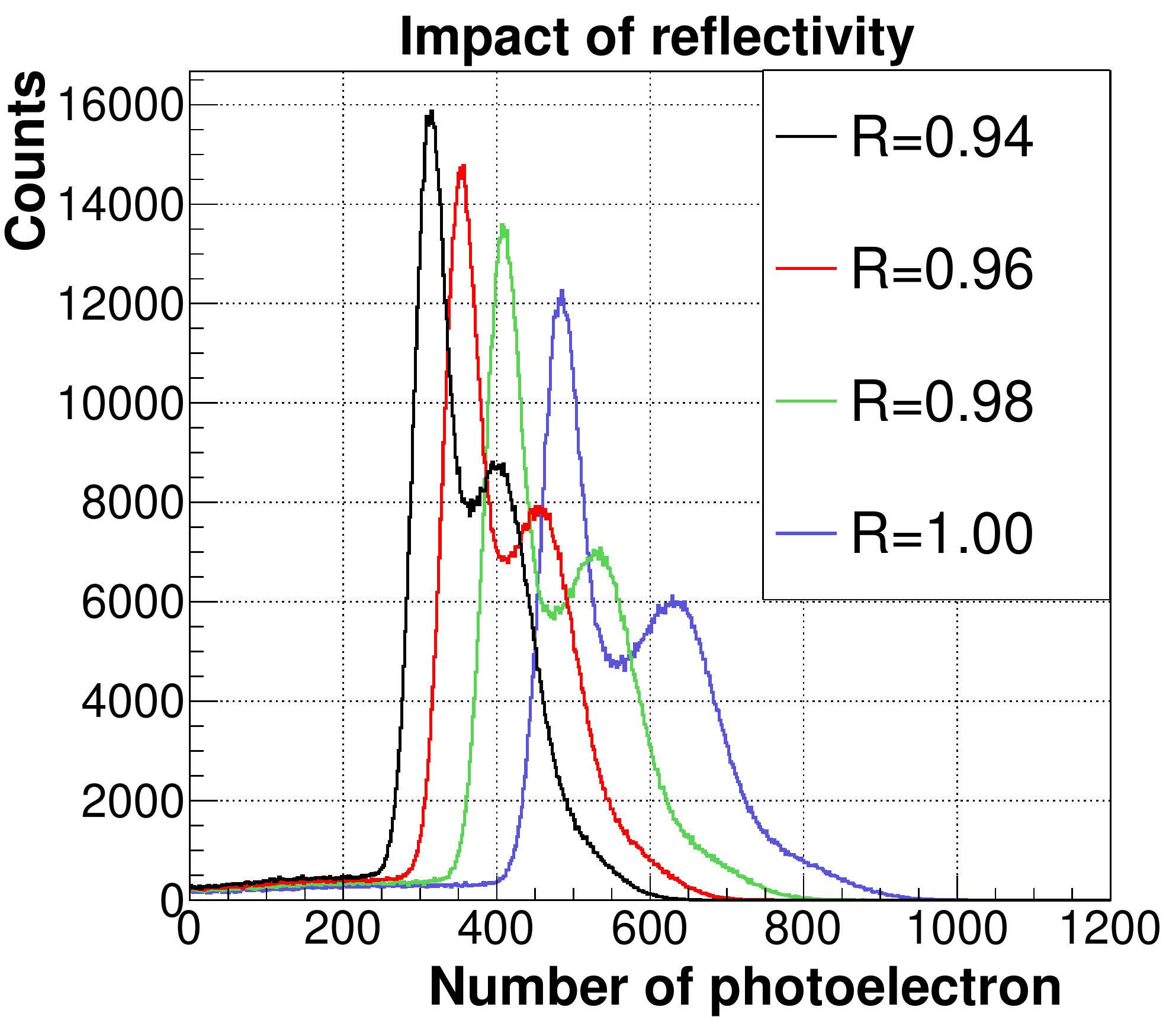}
        \label{fig:posref}
    }
    \\
    \subfigure[$^{60}Co$]
    {
       \includegraphics[width=0.47\linewidth]{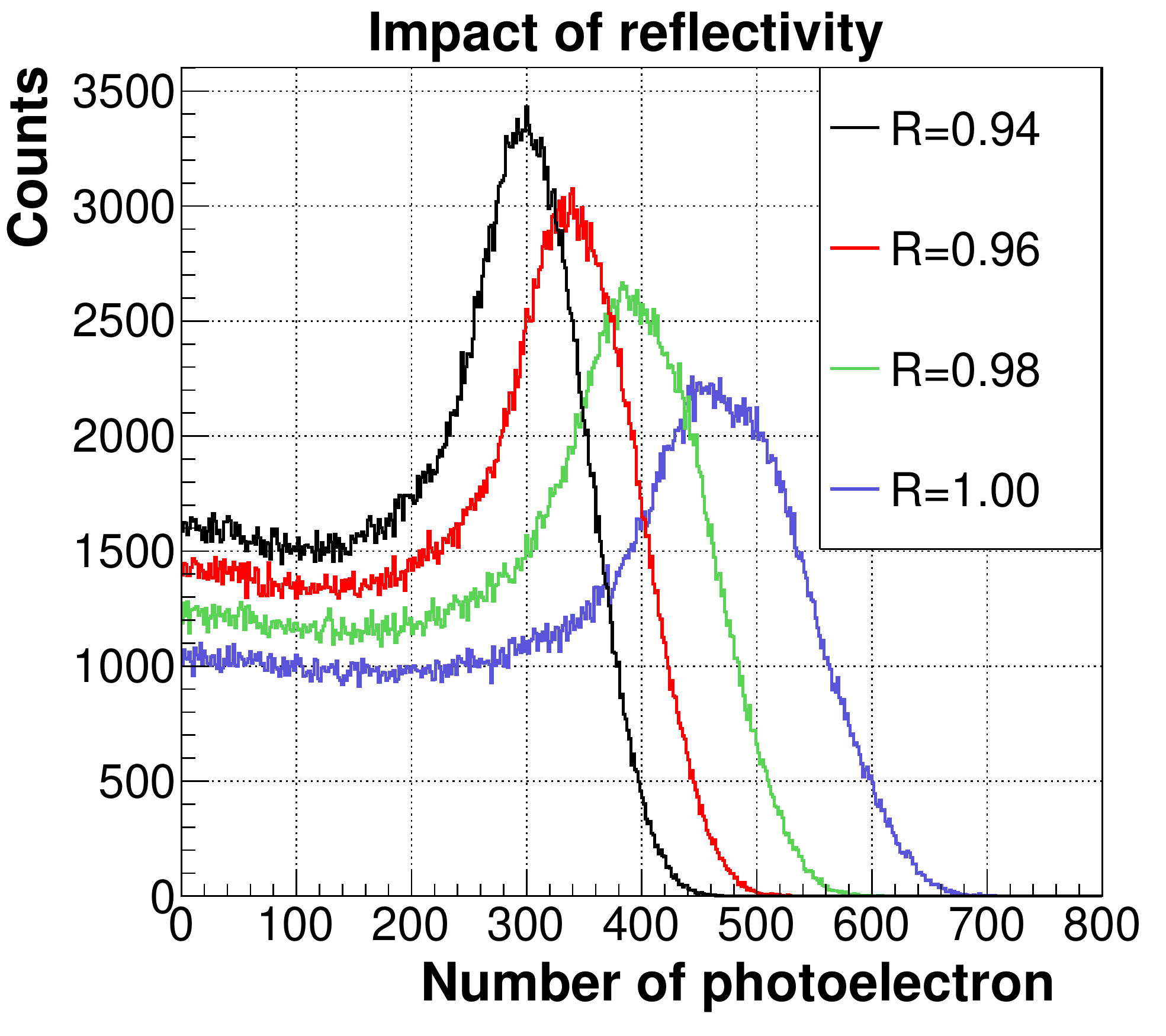}
        \label{fig:coref}
    }
    \subfigure[$^{137}Cs$]
    {
       \includegraphics[width=0.47\linewidth]{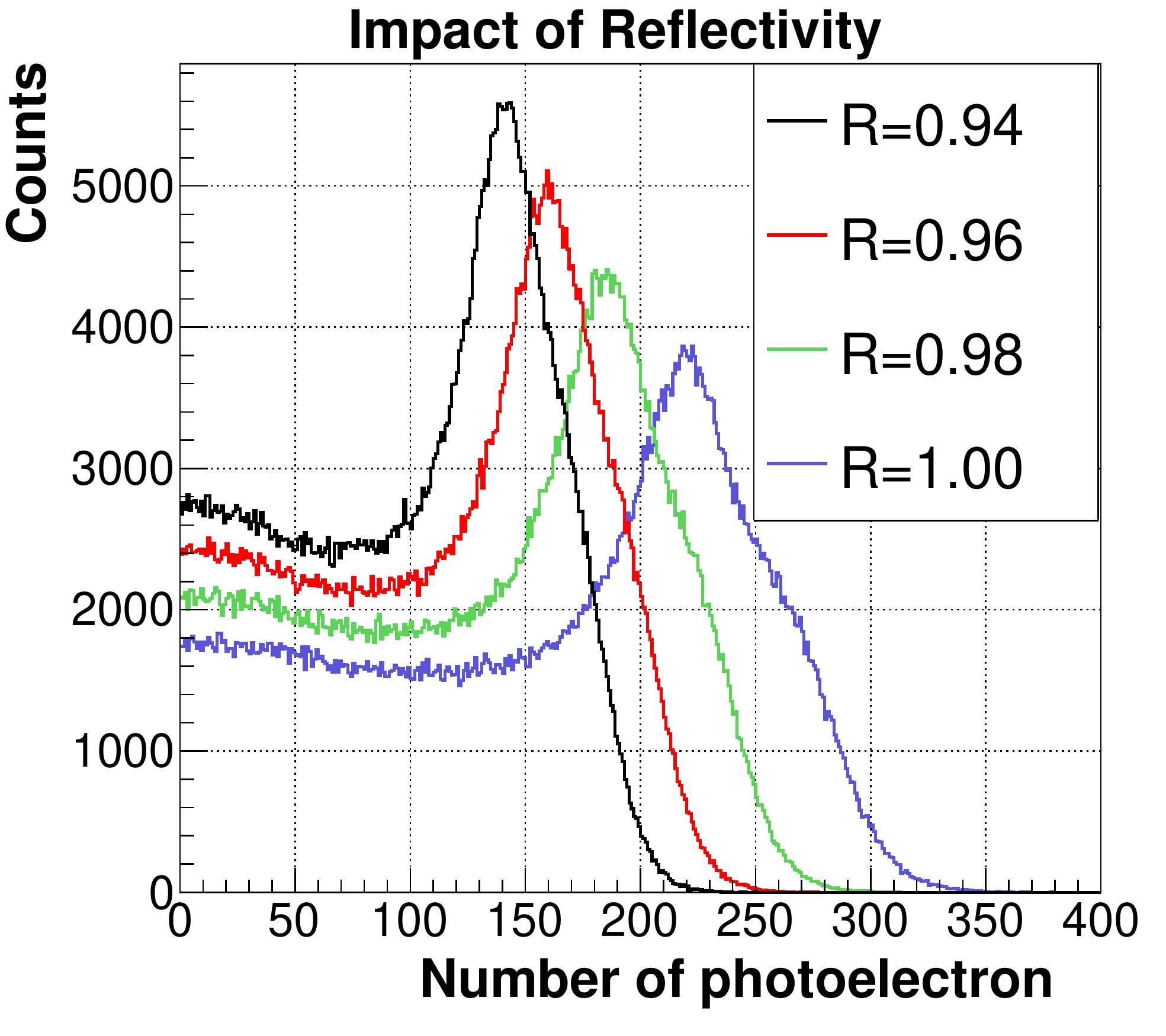}
        \label{fig:csref}
    }
    \caption{ The impact of reflectivity on photoelectron spectrum. Module 1 and 2-inch H6410 Hamamatsu PMT are used to perform simulation. The parameter $\sigma_{\alpha}$ set to 0.2.}
    \label{fig:REF}

\end{figure}

The influence of reflectivity on photoelectron spectra is investigated simulating the energy deposition of 1 MeV positron (Fig. $\ref{fig:posref}$), $^{60}Co$ (Fig. $\ref{fig:coref}$) and $^{137}Cs$ (Fig. $\ref{fig:csref}$). Fig. $\ref{fig:REF}$ compares the effect of reflecting material reflectivity value on detected photon spectrum when module 1 is wrapped slightly with a specular reflector (PBP). This figure shows that as the probability of reflection from reflective surface increases, the peak position of the spectrums shifts to the right. For Fig. $\ref{fig:posref}$, with respect to R=0.94, the peak position (first peak) shifts by $13\%$ (R=0.96), $30\%$ (R=0.98) and $55\%$ (R=1.0). Another important point is that the change in high reflectivity values shifts the spectrum further to the right. As a result, an increase from R=0.94 to R=0.96 shifts the spectrum $13\%$ to the right while an increase from R=0.98 to R=1.0 shifts the spectrum $18\%$ to the right. The simulation is also done for other surface finish type (GBP, PFP and GFP). The effect of reflectivity on detected spectrum provides similar spectrum.

\subsection{Impact of scintillator surface roughness on LCE and detected spectrum}
\label{sec:sigma} 

The effect of scintillator surface roughness degree (or scintillator surface polishing level) on light collection efficiency is investigated in the case of two different surface finish \footnote{The parameter $\sigma_{\alpha}$ is not defined for front painted method (PFP, GFP) since there is no air gap between the scintillator and the reflector.}: PBP and GBP (See Fig.$\ref{fig:Wrap}$). For both surface finish case, average light collection efficiency and its standard deviation are calculated for different $\sigma_{\alpha}$ value: 0.05, 0.10, 0.15 and 0.20. The results are shown in Fig. $\ref{fig:LCESA}$. This figure clearly shows that the best result is achieved when $\sigma_{\alpha}$=0.2 which correspond to a rough scintillator surface.\footnote{The parameter $\sigma_{\alpha}$=0.05 correspond to nearly polished surface while the parameter $\sigma_{\alpha}$=0.2 correspond to a ground surface \cite{Janecek2} } Because a rough scintillator surface partially prevents internal trapping of photons within the scintillator and consequently decreases self-absorption rate in the scintillator. The important point here is that the degrees of scintillator surface roughness. It should be optimum level (Here $\sigma_{\alpha}$=0.2). Increasing the surface roughness more than 0.2 gives us less LCE value which is not shown in Fig. $\ref{fig:LCESA}$.

\begin{figure}[!htb]
    \centering
    \subfigure[Positron]
    {
        \includegraphics[width=0.50\linewidth]{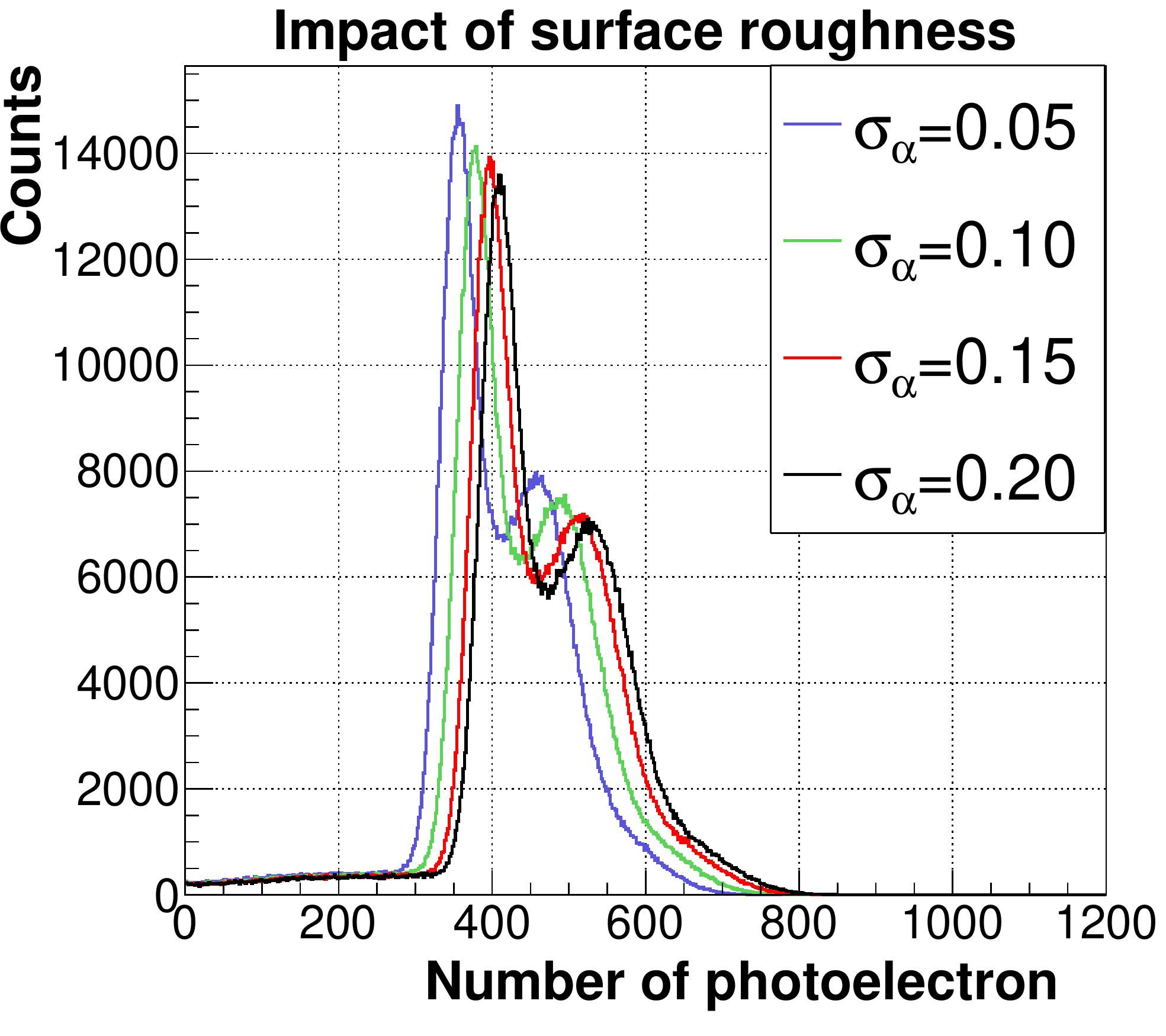}
        \label{fig:possa}
    }
    
    \subfigure[$^{60}Co$]
    {
       \includegraphics[width=0.47\linewidth]{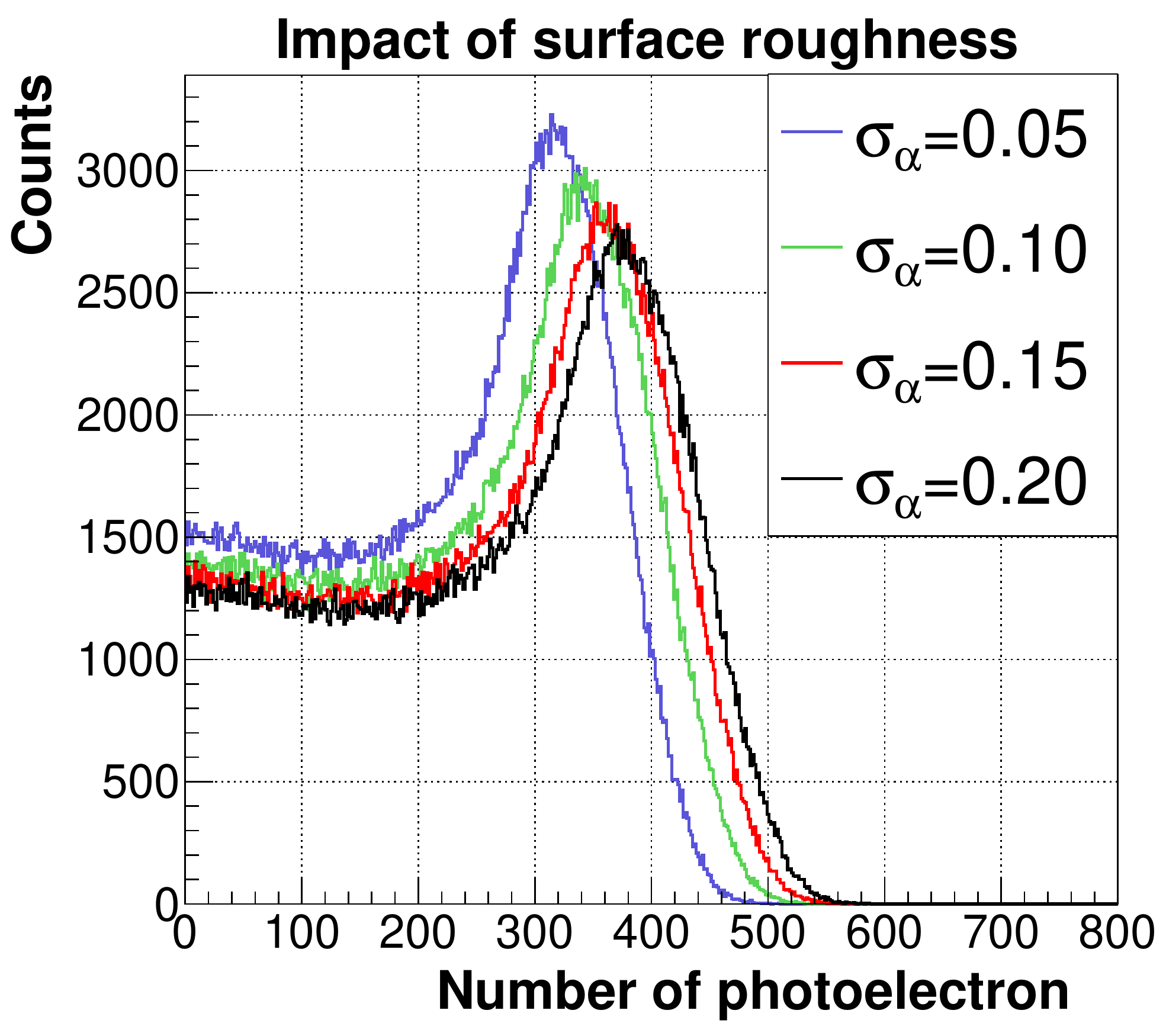}
        \label{fig:cosa}
    }
    \subfigure[$^{137}Cs$]
    {
       \includegraphics[width=0.47\linewidth]{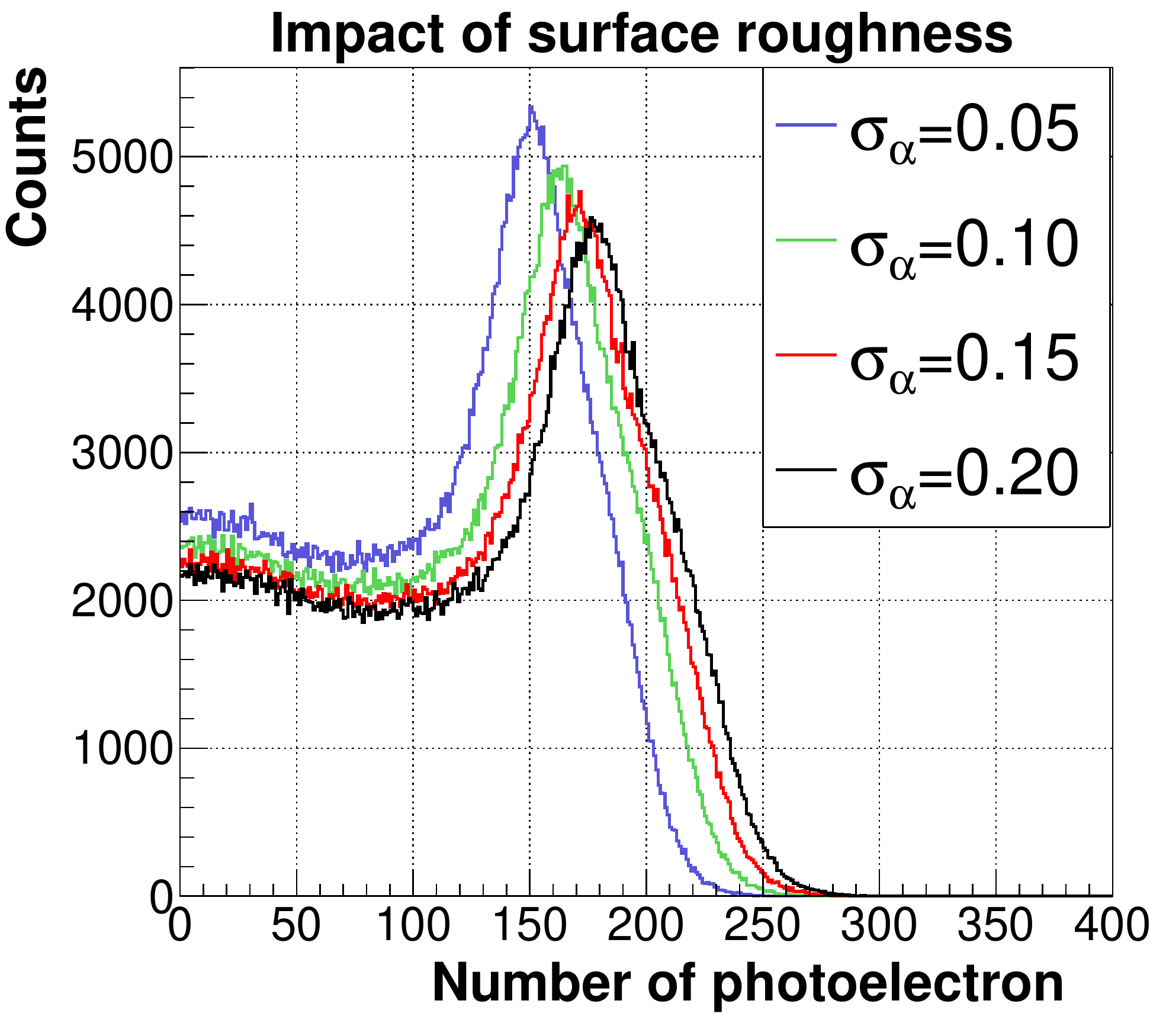}
        \label{fig:cssa}
    }
    \caption{The impact of scintillator surface polishing level on photoelectron spectra. Module 1 and 2-inch H6410 Hamamatsu PMT are used to perform simulation. Surface finish type set to PBP. The reflectivity of reflector set to R=0.98. The reflection type from scintillator-air interface set to specular lobe (SL=1.0)}
    \label{fig:SA}

\end{figure}

The effect of scintillator surface polishing level on photoelectron spectra is investigated simulating the energy deposition of 1 MeV positron (Fig. $\ref{fig:possa}$), $^{60}Co$ (Fig. $\ref{fig:cosa}$) and $^{137}Cs$ (Fig. $\ref{fig:cssa}$). Fig. $\ref{fig:SA}$ compares the effect of scintillator surface roughness on photoelectron spectra when module 1 is wrapped slightly with a specular reflector (PBP). As the parameter $\sigma_{\alpha}$ increases, the peak position of the spectrum shifts to the right. This shift is more noticeable in low value changes of $\sigma_{\alpha}$ . For Fig. $\ref{fig:possa}$, with respect to $\sigma_{\alpha}$=0.05, the peak position shifts by $6\%$ ($\sigma_{\alpha}$=0.10), $11\%$ ($\sigma_{\alpha}$=0.15), $15\%$ ($\sigma_{\alpha}$=0.20). Increasing $\sigma_{\alpha}$ more than 0.2 decreases the detected photon number and the peak position shifts to the left.

\begin{figure}[!htb]
    \centering
    \subfigure[Simulation parameters are set according to the experimental setup. The parameter $\sigma_{\alpha}$ indicates the degree of surface polishing. Three different reflectivity values are used in the simulation. Surface finish type set to PBP.  ]
    {
       \includegraphics[width=0.46\linewidth]{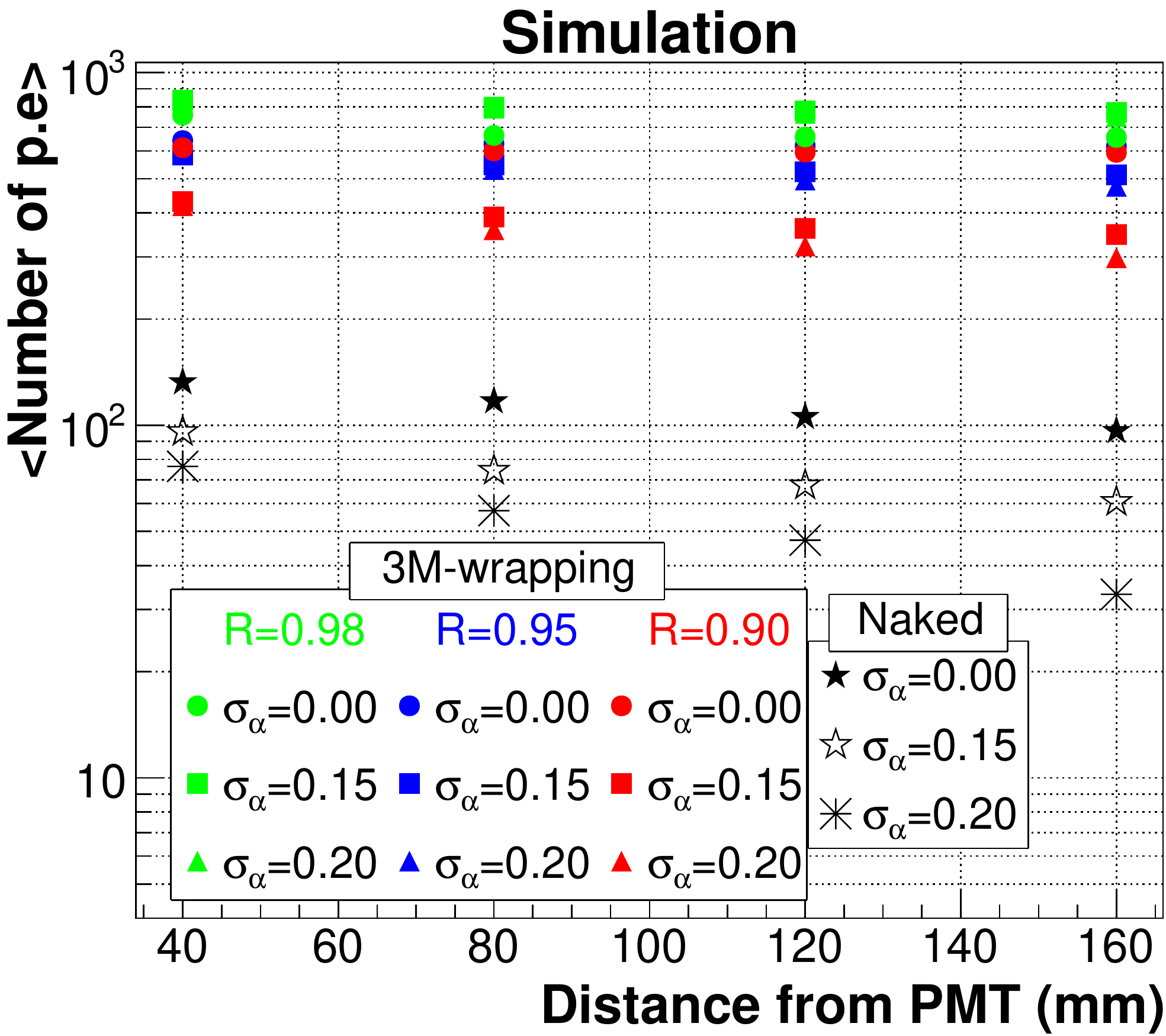}
        \label{fig:sim1}
    }
    \quad
    \subfigure[Experimental results are obtained from Ref. \cite{Gierlik}. The numbers denoted by symbols indicate the degree of polishing. The higher the number, the better the polishing. The abbreviation d.p stands for diamond polishing paste.]
    {
        \includegraphics[width=0.46\linewidth]{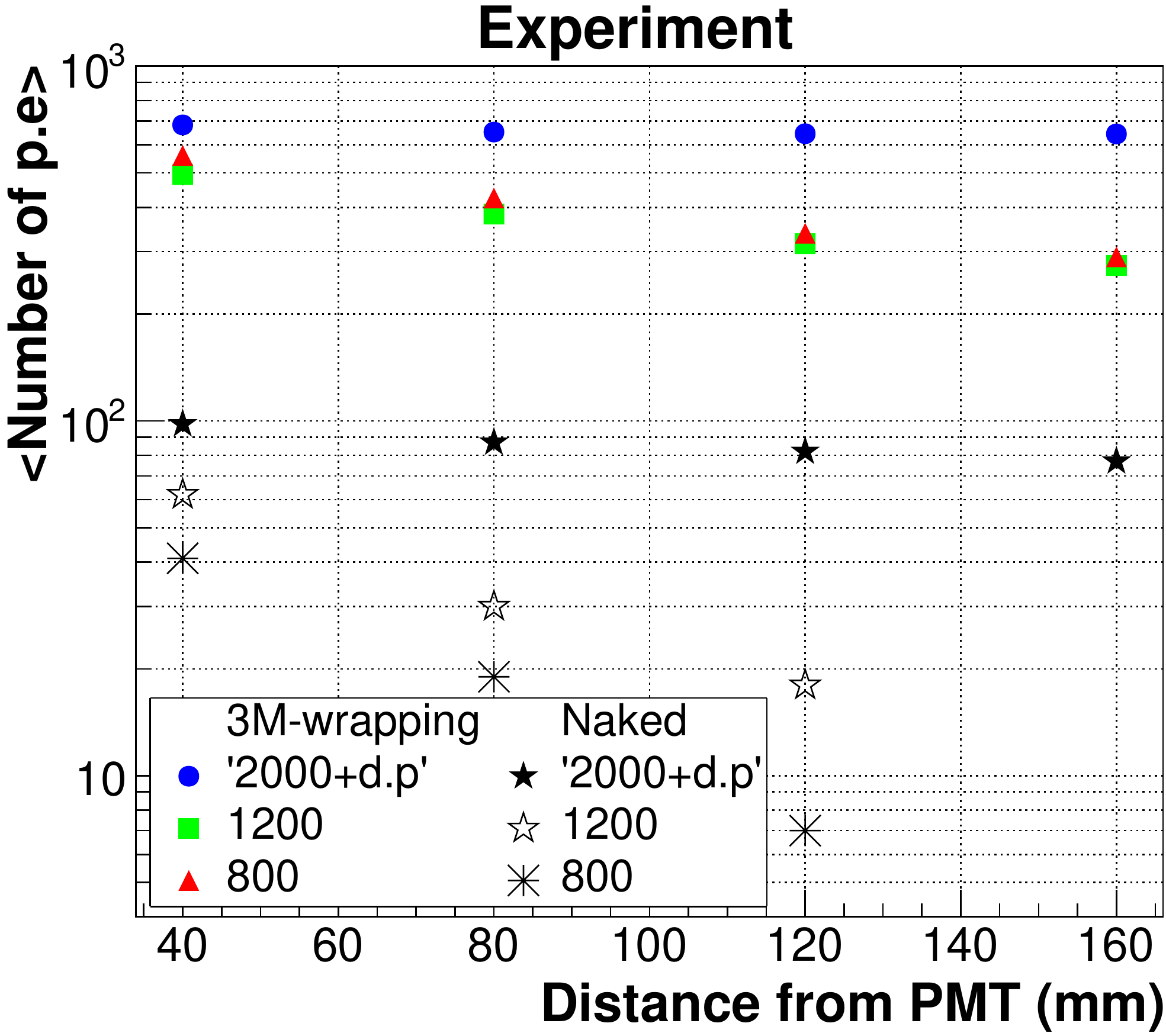}
        \label{fig:exp1}
    }
     \caption{A comparison between the simulation and experiment for three different degrees of polishing. The parameters $\sigma_{\alpha}=0.00$, $\sigma_{\alpha}=0.15$, $\sigma_{\alpha}=0.20$ in the simulations correspond to the parameters '2000 + d.p', '1200', '800' in the experiment, respectively. Since the exact match is not estimated, the points in the simulation and the experiment may show different value. Here, the important point is the general impact of surface polishing on light collection. $^{137}Cs$ gamma source is used in the experiment to generate optical photons. Different distances from source to PMT are considered in the experiment. }
    \label{fig:comp1}
    
\end{figure}

In order to assess the validity of the simulation, it is compared with an experiment which shows the influence of surface polishing on light collection efficiency by applying three different degrees of polishing \footnote{Grinder granularity of '800', '1200', and '2000 + diamond paste (d.p)' is used in the experiment to specify the degrees of polishing quantitatively (the higher the number, the better the polishing).} onto the surface of 6mm$\times$6mm $\times$200mm EJ200 plastic scintillator bar \cite{Gierlik}. For an exact comparison, the simulation parameters are set according to the experimental setup. Fig. $\ref{fig:comp1}$ shows the results of the experiment and simulation. In the case of using a naked scintillator, the simulation and experiment show the similar behavior: The better the surface polishing, the higher the light collection. On the other hand, in the case of using a wrapped scintillator with the specular reflector (3M Enhanced Specular Reflector is used in the experiment which corresponds to PBP parameter in the simulation), the simulation reveals that there is a critical reflectivity value for each wrapped scintillator geometry which changes the impact of surface polishing on LCE. This critical value is determined to be 0.97 for the geometry used in the experiment. Below this value, LCE increases as the degree of polishing increases (the case of R=0.90 and R=0.95 in Fig. $\ref{fig:sim1}$). If the reflectivity of the reflector exceeds this critical value, a rough scintillator surface becomes more favorable for light collection (the case of R=0.98 in Fig. $\ref{fig:sim1}$). As in the simulation, the highest light collection is obtained with the case of '2000+d.p' in the experiment. However, the case of '800' collects very little light than the case of '1200', which is reverse in the simulation.\footnote{ This little difference may arise from the reflector changing the degree of polishing.} Finally, with regard to our actual scintillator geometry (antineutrino detector modules), a scan is performed over all reflectivity values and the critical value is found to be 0.9.

%\clearpage
 
\section{Conclusion}

In this study, many factors affecting the light collection efficiency of antineutrino detector module have been studied. Firstly, the effect of light guide shape and photocathode size on light collection efficiency are investigated. It has been found that the highest light collection efficiency is achieved when module 2 is used. With the $27\%$ efficiency of light collection, module 2 is seen to collect $8\%$ more light than module 1 and 6$\%$ more light than module 3 when 4.6cm diameter photocathode size (the size of Hamamatsu PMT photocathode) is used. A significant increase in light collection efficiency is arised from the size of the photocathode. When the radius of the photocathode is increased from 2.3cm (2-inch H6410 Hamamatsu PMT) to 3.5cm (3-inch 9265B Et Enterprise PMT), the light collection efficiency increases by $89\%$, $51\%$ and $76\%$ for module 1, module 2 and module 3 respectively. Secondly, the most efficient reflector type and its applying method onto scintillator and light guide surface are determined. The results show that wrapping the scintillator and light guide loosely with a specular reflector (PBP) gives the highest light collection efficiency. Thirdly, the impact of reflectivity of reflecting material on light collection is examined. It is observed that small changes in reflection probability of the reflector significantly change the efficiency of light collection especially at high reflectivity value. For instance, a drop from R=0.98 to R=0.96 reduces the light collection efficiency by $15\%$ for PBP surface finish case. Finally, the influence of scintillator surface polishing on light collection efficiency is explored. The critical reflectivity value for antineutrino detector modules is determined to be 0.9. Since the reflectivity values used in the simulation is greater than this critical value, a rough scintillator surface is selected for better light collection. The optimal value of scintillator surface roughness degree is searched. It is found that a ground scintillator surface with the value of $\sigma_{\alpha}=0.2$ exhibits the best performance in terms of light collection. The obtained simulation results show consistent outcomes with the experimental findings. An optimized antineutrino detector module according to these parameters provides a detector with a higher energy resolution and consequently improves sensitivity of the detector to the changes of fission isotopes of the reactor core.

\clearpage

%\section{Section in Appendix}
%\label{appendix-sec1}

%% References


\bibliographystyle{elsarticle-num}

\bibliography{sample2}

\begin{thebibliography}{10}
\expandafter\ifx\csname url\endcsname\relax
  \def\url#1{\texttt{#1}}\fi
\expandafter\ifx\csname urlprefix\endcsname\relax\def\urlprefix{URL }\fi
\expandafter\ifx\csname href\endcsname\relax
  \def\href#1#2{#2} \def\path#1{#1}\fi

\bibitem{Mikaelyan}
L.~A. {Mikaelyan}, {Neutrino Laboratory in the Atomic Plant (Fundamental and
  Applied Researches)}, in: Neutrino 77, Volume 2, 1978, p. 383.

\bibitem{Yuvv}
Y.~V. Klimov, V.~I. Kopeikin, L.~A. Mikaélyan, K.~V. Ozerov, V.~V. Sinev,
  Neutrino method remote measurement of reactor power and power output, Atomic
  Energy 76~(2) (1994) 123--127.
\newblock \href {http://dx.doi.org/10.1007/bf02414355}
  {\path{doi:10.1007/bf02414355}}.

\bibitem{Bernstein}
A.~Bernstein, N.~S. Bowden, A.~Misner, T.~Palmer, Monitoring the thermal power
  of nuclear reactors with a prototype cubic meter antineutrino detector,
  Journal of Applied Physics 103~(7) (2008) 074905.
\newblock \href {http://dx.doi.org/10.1063/1.2899178}
  {\path{doi:10.1063/1.2899178}}.

\bibitem{Pequignot}
M.~Pequignot, The nucifer and stereo reactor antineutrino experiments, Nuclear
  and Particle Physics Proceedings 265-266 (2015) 126–128.
\newblock \href {http://dx.doi.org/10.1016/j.nuclphysbps.2015.06.032}
  {\path{doi:10.1016/j.nuclphysbps.2015.06.032}}.

\bibitem{Anjos}
J.~Anjos, T.~Abrahão, T.~Alvarenga, L.~Andrade, G.~Azzi, A.~Cerqueira,
  P.~Chimenti, J.~Costa, T.~Dornelas, P.~Farias, et~al., Using neutrinos to
  monitor nuclear reactors: the angra neutrino experiment, simulation and
  detector status, Nuclear and Particle Physics Proceedings 267-269 (2015)
  108–115.
\newblock \href {http://dx.doi.org/10.1016/j.nuclphysbps.2015.10.090}
  {\path{doi:10.1016/j.nuclphysbps.2015.10.090}}.

\bibitem{ABernstein}
A.~Bernstein, G.~Baldwin, B.~Boyer, M.~Goodman, J.~Learned, J.~Lund, D.~Reyna,
  R.~Svoboda, Nuclear security applications of antineutrino detectors: Current
  capabilities and future prospects, Science \& Global Security 18~(3) (2010)
  127–192.
\newblock \href {http://dx.doi.org/10.1080/08929882.2010.529785}
  {\path{doi:10.1080/08929882.2010.529785}}.

\bibitem{Oguri}
S.~Oguri, Y.~Kuroda, Y.~Kato, R.~Nakata, Y.~Inoue, C.~Ito, M.~Minowa, Reactor
  antineutrino monitoring with a plastic scintillator array as a new safeguards
  method, Nuclear Instruments and Methods in Physics Research Section A:
  Accelerators, Spectrometers, Detectors and Associated Equipment 757 (2014)
  33–39.
\newblock \href {http://dx.doi.org/10.1016/j.nima.2014.04.065}
  {\path{doi:10.1016/j.nima.2014.04.065}}.

\bibitem{Battaglieri}
M.~Battaglieri, R.~Devita, G.~Firpo, P.~Neuhold, M.~Osipenko, D.~Piombo,
  G.~Ricco, M.~Ripani, M.~Taiuti, An anti-neutrino detector to monitor nuclear
  reactors power and fuel composition, Nuclear Instruments and Methods in
  Physics Research Section A: Accelerators, Spectrometers, Detectors and
  Associated Equipment 617~(1-3) (2010) 209–213.
\newblock \href {http://dx.doi.org/10.1016/j.nima.2009.09.031}
  {\path{doi:10.1016/j.nima.2009.09.031}}.

\bibitem{Alekseev}
I.~Alekseev, V.~Belov, V.~Brudanin, M.~Danilov, V.~Egorov, D.~Filosofov,
  M.~Fomina, Z.~Hons, S.~Kazartsev, A.~Kobyakin, et~al., Detector of the
  reactor antineutrino based on solid-state plastic scintillator (danss).
  status and first results., Journal of Physics: Conference Series 798 (2017)
  012152.
\newblock \href {http://dx.doi.org/10.1088/1742-6596/798/1/012152}
  {\path{doi:10.1088/1742-6596/798/1/012152}}.

\bibitem{Georgadze}
A.~S. Georgadze, V.~M. Pavlovych, O.~A. Ponkratenko, D.~A. Litvinov, {A remote
  reactor monitoring with plastic scintillation detector}\href
  {http://arxiv.org/abs/1610.05884} {\path{arXiv:1610.05884}}.

\bibitem{Agostinelli}
S.~Agostinelli, J.~Allison, K.~Amako, J.~Apostolakis, et~al., Geant4—a
  simulation toolkit, Nuclear Instruments and Methods in Physics Research
  Section A: Accelerators, Spectrometers, Detectors and Associated Equipment
  506~(3) (2003) 250 -- 303.
\newblock \href {http://dx.doi.org/10.1016/S0168-9002(03)01368-8}
  {\path{doi:10.1016/S0168-9002(03)01368-8}}.

\bibitem{Levin}
A.~Levin, C.~Moisan, A more physical approach to model the surface treatment of
  scintillation counters and its implementation into detect, 1996 IEEE Nuclear
  Science Symposium. Conference Record\href
  {http://dx.doi.org/10.1109/nssmic.1996.591410}
  {\path{doi:10.1109/nssmic.1996.591410}}.

\bibitem{Eljen}
{Eljen Technology}, {EJ-200 Plastic Scintillator, EJ-500 Optical Cement, EJ-560
  Slicon Rubber}, \url{http://www.eljentechnology.com/products}.

\bibitem{Hamamatsu}
{Hamamatsu Photonics}, {60 mm (2") Photomultiplier tube assembly H6410},
  \url{https://www.hamamatsu.com/us/en/product/alpha/P/3002/H6410/index.html}.

\bibitem{Et}
{ET Enterprises Limited}, {78 mm (3") photomultiplier 9265B series data sheet},
  \url{https://my.et-enterprises.com/pdf/9265B.pdf}.

\bibitem{taheri}
A.~Taheri, R.~G. Peyvandi, The impact of wrapping method and reflector type on
  the performance of rod plastic scintillators, Measurement 97 (2017)
  100–110.
\newblock \href {http://dx.doi.org/10.1016/j.measurement.2016.10.051}
  {\path{doi:10.1016/j.measurement.2016.10.051}}.

\bibitem{Janecek}
M.~Janecek, Reflectivity spectra for commonly used reflectors, IEEE
  Transactions on Nuclear Science 59~(3) (2012) 490–497.
\newblock \href {http://dx.doi.org/10.1109/tns.2012.2183385}
  {\path{doi:10.1109/tns.2012.2183385}}.

\bibitem{Janecek2}
M.~Janecek, W.~W. Moses, Simulating scintillator light collection using
  measured optical reflectance, IEEE Transactions on Nuclear Science 57~(3)
  (2010) 964–970.
\newblock \href {http://dx.doi.org/10.1109/tns.2010.2042731}
  {\path{doi:10.1109/tns.2010.2042731}}.

\bibitem{Gierlik}
M.~Gierlik, T.~Batsch, R.~Marcinkowski, M.~Moszyński, T.~Sworobowicz, Light
  transport in long, plastic scintillators, Nuclear Instruments and Methods in
  Physics Research Section A: Accelerators, Spectrometers, Detectors and
  Associated Equipment 593~(3) (2008) 426–430.
\newblock \href {http://dx.doi.org/10.1016/j.nima.2008.05.030}
  {\path{doi:10.1016/j.nima.2008.05.030}}.

\end{thebibliography}

\end{document}